\documentclass[aps, prd, twocolumn, amsmath, floats,floatfix, superscriptaddress, nofootinbib, showkeys]{revtex4}

\usepackage{amssymb}
\usepackage{amsmath}
\usepackage{verbatim}
\usepackage{mathrsfs}
\usepackage{amsfonts}
\usepackage{amsfonts} 
\usepackage{latexsym}
\usepackage{epsfig}
\usepackage{color}
\usepackage{graphicx,subfigure}
\usepackage{units}
\usepackage{envmath}
\usepackage{natbib}
\usepackage{ctable}
\usepackage{soul}
\usepackage{ulem}
\usepackage{bigints}

\begin{document}
	
	
	\definecolor{orange}{rgb}{0.9,0.45,0}
	
	\newcommand{\re}{\mbox{Re}}
	\newcommand{\im}{\mbox{Im}}

	\def\CovDev{D}
	\def\Res{{\mathcal R}}
	\def\Gammaflat{\hat \Gamma}
	\def\metricflat{\hat \gamma}
	\def\Dflat{\hat {\mathcal D}}
	\def\part_n{\partial_\perp}
	
	\def\Lie{\mathcal{L}}
	\def\A{\mathcal{X}}
	\def\Aphi{\A_{\phi}}
	\def\hAphi{\hat{\A}_{\phi}}
	\def\E{\mathcal{E}}
	\def\Ham{\mathcal{H}}
	\def\M{\mathcal{M}}
	\def\R{\mathcal{R}}
	\def\p{\partial}
	
	\def\hg{\hat{\gamma}}
	\def\hA{\hat{A}}
	\def\hD{\hat{D}}
	\def\hE{\hat{E}}
	\def\hR{\hat{R}}
	\def\hcA{\hat{\mathcal{A}}}
	\def\hDelt{\hat{\triangle}}
	
	\def\na{\nabla}
	\def\dif{{\rm{d}}}
	\def\non{\nonumber}
	\newcommand{\erf}{\textrm{erf}}
	
	\renewcommand{\t}{\times}
	
	\long\def\symbolfootnote[#1]#2{\begingroup%
		\def\thefootnote{\fnsymbol{footnote}}\footnote[#1]{#2}\endgroup}

	
\title{The Integrated Sachs-Wolfe Effect in Interacting Dark Matter-Dark Energy Models}
\author{Mina Ghodsi Yengejeh}
\email{m.ghodsi.y@gmail.com}
\affiliation{PDAT Laboratory, Department of Physics, K.N. Toosi University of Technology, P.O. Box 15875-4416, Tehran, Iran}
\affiliation{Department of Physics, Shahid Beheshti University, G.C., Evin, Tehran 19839, Iran}
\author{Saeed Fakhry}
\email{s_fakhry@sbu.ac.ir}
\affiliation{Department of Physics, Shahid Beheshti University, G.C., Evin, Tehran 19839, Iran}
\author{Javad T. Firouzjaee}
\email{firouzjaee@kntu.ac.ir}
\affiliation{Department of Physics, K.N. Toosi University of Technology, P.O. Box 15875-4416, Tehran, Iran}
\affiliation{School of Physics, Institute for Research in Fundamental Sciences (IPM), P.O. Box 19395-5531, Tehran, Iran}
\affiliation{PDAT Laboratory, Department of Physics, K.N. Toosi University of Technology, P.O. Box 15875-4416, Tehran, Iran}
\author{Hojatollah Fathi}
\email{h.fathi1998@gmail.com}
\affiliation{PDAT Laboratory, Department of Physics, K.N. Toosi University of Technology, P.O. Box 15875-4416, Tehran, Iran}
	
\date{\today}
	
	
\begin{abstract} 
	
Interacting dark matter-dark energy (IDMDE) models can be taken to account as one of the present challenges that may affect the cosmic structures. The main idea of such models is that the mass of dark matter particles can be specified by their interactions with a scalar field whose energy density is characterized by dark energy. In this work, we propose to study the integrated Sachs-Wolfe (ISW) effect in IDMDE models. To this end, we initially introduce a theoretical framework for IDMDE models. Moreover, we briefly discuss the stability conditions of IDMDE models and by specifying a simple functional form for the energy density transfer rate, we calculate the perturbation equations. In the following, we calculate the amplitude of the matter power spectrum for the IDMDE model and compare it with the corresponding result obtained from the $\Lambda$CDM model. Furthermore, we calculate the amplitude of the ISW auto-power spectrum as a function of multipole order $l$ for the IDMDE model. The results indicate that the amplitude of the ISW auto-power spectrum in the IDMDE model for different phantom dark energy equations of state behaves similar to the one for the $\Lambda$CDM model, whereas, for the quintessence dark energy equations of state, the amplitude of the ISW-auto power spectrum for the IDMDE model should be higher than the one for the $\Lambda$CDM model. Also, it turns out that the corresponding results by different values of the coupling parameter demonstrate that $\xi$ is inversely proportional to the amplitude of the ISW-auto power spectrum in the IDMDE model. Finally, by employing four different surveys, we calculate the amplitude of the ISW-cross power spectrum as a function of multipole order $l$ for the IDMDE model. The results exhibit that the amplitude of the ISW-cross power spectrum for the IDMDE model for all values of $\omega_{\rm x}$ is higher than the one obtained for the $\Lambda$CDM model, while deviations are still less than $0.1$ order of magnitude. Also, it turns out that the amplitude of the ISW-cross power spectrum in the IDMDE model changes inversely with the value of coupling parameter $\xi$.	
\end{abstract}
	
	\maketitle
	
	\vspace{0.8cm}

	\section{Introduction}\label{sec:intro}
The standard model of cosmology, the Lambda cold dark matter ($\Lambda$CDM) model, gained big success in explaining wide observations in large-scale physics. This success is due to the great efforts to set up cosmological surveys like the cosmic microwave background (CMB) \cite{WMAP:2003gmp,WMAP:2006bqn,Planck:2015fie,Planck:2018vyg}, type Ia supernovae \cite{SupernovaSearchTeam:1998fmf, SupernovaSearchTeam:1998fmf2} , galaxy surveys \cite{SDSS1, SDSS2}, and weak lensing \cite{Hildebrandt:2016iqg,DES:2017myr}. All these surveys have confirmed the $\Lambda$CDM \cite{Peebles}, which shows the entire matter budget of the Universe consists primarily of cold dark matter and dark energy in the present-time Universe. 

Recently, some serious tensions in observational data have opened some new windows to think about the nature of dark energy and dark matter beyond the $\Lambda$CDM model. The most impressive examples of so-called tensions between the $\Lambda$CDM predictions can be deduced from the latest analysis of the Planck collaboration \cite{Planck:2018vyg} and independent observational of the distance ladder measurements. These measurements employ close cepheids as anchors for measuring the Hubble parameter $H_{0}$ that measures the expansion rate of the Universe.

Hubble space telescope (HST) has measured this tension through local distance ladder measurements, and the supernova $H_{0}$ for the equation of state (SH0ES) collaboration reports $H_{\rm 0} = 74.03 \pm 1.42\,{\rm km\,s^{-1}\,Mpc^{-1}}$ \cite{Riess:2019cxk}, which is at a $4.4 \sigma$ difference to the value measured by the Planck satellite, i.e., $H_{\rm 0} = 67.4 \pm 0.5\,{\rm km\,s^{-1}\,Mpc^{-1}}$ \cite{Planck:2018vyg}. Another tension is the value of $\sigma_{\rm 8}$, which represents the amount of late-time matter clustering. To this end, some surveys such as the kilo degree survey (KiDS) \cite{Kohlinger:2017sxk} and the dark energy survey (DES) \cite{DES:2017myr, DES:2017qwj, DES:2017qwj2} have been utilized, resulting in tensions in $\sigma_{8}$ measurements of the order of $2\sim 3\,\sigma$.

Many different studies have been performed to solve these tensions \cite{H0review}, including models that question the nature of dark matter or dark energy, which can be studied in mass-temperature relation \cite{Naseri:2020uvn} and scaling relation \cite{Naseri:2020dgq} in the galaxy clusters. Moreover, various cosmological aspects have been discussed, while considering other dark matter candidates \cite{Bernal:2018ins, Hryczuk:2020jhi, Egana-Ugrinovic:2021gnu, Fakhry:2020plg, Fakhry:2021tzk}. A model that has been studied extensively is the interaction between dark matter and dark energy (henceforth IDMDE), see e.g., \cite{Wang:2016lxa} for a recent review.

Another interesting approach to constrain dark energy models is to investigate the integrated Sachs-Wolfe (ISW) effect in the observational data in such a way that the cross-correlation between the ISW signal and the galaxy distribution can be specified. The ISW effect is an additional effect associated with the time-dependence gravitational potential as photons pass through the observable Universe. It should be noted that gravitational potentials are constant in time during the matter-dominated era. Hence, emitted photons from the last scattering surface do not change during the matter-dominated era. Except at very large scales, the amplitude of the ISW fluctuations tends to be smaller than that of fluctuations originating at the epoch of the last scattering. However, the CMB fluctuations were created much more recently. Therefore, they are expected to be partially correlated with tracers of the large-scale matter distribution, such as distant galaxies. In this way, the ISW effect results in a new direction in observational cosmology to constrain cosmological models with time-varying potentials.

Since the ISW affects the redshift of the measurement as an integrated effect on large redshift ranges (from the last scattering surface to now) can be more sensitive to variations in dark energy properties. In this regard, calculation of the ISW effect can be a convenient criterion for studying dark energy \cite{Crittenden:1995ak} and can probe some modifications of the $\Lambda$CDM model such as interacting dark sector models and modified gravity \cite{Munshi:2014tua, Adamek:2019vko, Kable:2021yws, Y:2021ybx}.

In this work, we calculate the ISW effect within the framework of an IDMDE model. In this respect, the outline of this work is structured as below. In Sec.~\ref{sec:ii}, we describe the theoretical framework of IDMDE models. In this regard, by considering a perturbed metric for the inhomogeneous Universe, we determine density and velocity perturbation equations. Then in Sec.~\ref{sec:iii}, we briefly discuss the stability conditions of IDMDE models and rewrite the perturbation equations by determining a simple functional form for the energy density transfer rate. Furthermore, in Sec.~\ref{sec:iv}, we calculate the ISW effect within the context of IDMDE models and compare our results with that of the $\Lambda$CDM model. Finally, we scrutinize the results and summarize the findings in Sec.~\ref{sec:iv}.
	\section{Interacting Dark Matter-Dark Energy Models}\label{sec:ii}
	Astrophysical and cosmological experiments such as the measurement of temperature anisotropies from the CMB \cite{WMAP:2006bqn}, large-scale structure evaluations \cite{2DFGRS:2001zay, SDSS:2003eyi, 2dFGRS:2005yhx, Springel:2006vs}, the integrated Sachs-Wolfe effect \cite{Fosalba:2003ge, boughn2004, Boughn:2004zm, boughn2005}, luminosity distances estimation via supernovae type Ia \cite{Boomerang:2000efg, SupernovaCosmologyProject:2003dcn, SupernovaSearchTeam:2003cyd, SupernovaSearchTeam:2004lze, SNLS:2005qlf}, and weak lensing predictions \cite{Contaldi:2003hi} demonstrate that the late-time Universe is passing an accelerated expansion phase. In this regard, the standard model of cosmology suggests an unknown relativistic matter component with negative pressure, referred to as dark energy, to explain the accelerated expansion phase in the late-time Universe. It is believed that dark energy is currently about $75\%$ of the total density of the Universe and the remaining $25\%$ is specified by the density of nonrelativistic matter component (i.e., dark matter and baryon) and very small amounts of relativistic particles (i.e., neutrinos and photons). Thus, the dynamics of the accelerated expansion phase may be referred to as a theoretical paradigm called the cosmological constant. However, providing a cosmological constant to justify the accelerated expansion phase in the late-time Universe, despite all advantages, would not be perfect. In other words, the discrepancy between observed values of cosmological constant and theoretical predictions of zero-point energy leads to an issue known as the cosmological constant problem \cite{wienberg1989a}. Additionally, observational experiments suggest that in the present-time Universe the ratio of the dark energy density to the matter one is of the order of unity, which requires special conditions to be satisfied in the early Universe. The question that may arise is why now $\rho_{\rm DM}/\rho_{\rm DE}\sim \mathcal{O}(1)$?, which is known as the coincidence problem \cite{Steinhardt1997}.
	
In order to address these problems, one can consider an interaction between the components of the dark sectors (i.e., matter and energy) in such a way that the ratio $\rho_{\rm DM}/\rho_{\rm DE}$ includes an evolution from a constant and unstable value in the early Universe to a lower constant and stable value in the late-time Universe and well explain the accelerated expansion era \cite{Zimdahl:2001ar, Chimento:2003iea}. In this regard, dark matter and dark energy densities do not evolve independently. On the other hand, one assumes that photon and baryon can not be coupled to dark energy. Hence, they do not contribute to interactions. Also, radiation can be neglected, as the focus is on the formation of structures in the late-time Universe \cite{Caldera-Cabral:2009hoy}.
	
Let us start with the Bianchi identity, $\nabla_{\rm \nu}G^{\mu \nu}=0$, which implies the energy-momentum conservation within the framework of the standard model of cosmolgy. While it leads to an interacting term in the continuity equations of dark energy and dark matter fluids in interacting models. In such a scenario, dark sector energy-momentum tensors cannot be conserved:
\begin{equation}
\nabla_{\rm \mu} T^{\rm \mu \nu}_{\rm A} = Q^{\rm \nu}_{\rm A},
\label{EMT-not-conserved}
\end{equation}
where $Q^{\rm \nu}$ is the four vector energy-momentum transfer rate, and subscript ``A" is allowed to be ``x" (for dark energy) and ``c" (for dark matter). It is essential to remind that the energy-momentum tensor of the total dark sector is conserved, namely
\begin{equation}
\sum_{\rm A} \nabla_{\rm \mu} T^{\rm \mu \nu}_{\rm A} = \sum_{\rm A} Q^{\rm \nu}_{\rm A} = 0,
\label{mod-cons-eq}
\end{equation}
which implies $Q_{\rm c}^{\rm \nu} = - Q_{\rm x}^{\rm \nu}$.
	
In order to investigate the impact of interacting models at large scales, one can assume a spatially flat and expanding Universe governed by a perfect fluid. In light of such considerations, the evolution of the energy-momentum density is determined by taking $\nu = 0$. Therefore, one can have the following energy balance equations for dark matter and dark energy
	\begin{equation}
	\rho \sp \prime_{\rm c} + 3 \mathcal{H} \rho_{\rm c} = - a^2 Q^0 = - a Q,
	\label{EEMDc}
	\end{equation}
	\begin{equation}
	\rho \sp \prime_{\rm x} + 3 \mathcal{H} (1+\omega_{\rm x}) \rho_{\rm x} = a^2 Q^0 = a Q,
	\label{EEMDx}
	\end{equation}
where prime denotes derivative with respect to the conformal time $\tau$, $a(\tau)$ is the scale factor, $Q\equiv Q^0/a$ is a coupling parameter called the energy density transfer rate, $\mathcal{H} = a \sp \prime/a$ is the conformal Hubble parameter and $\omega_{\rm x} = p_{\rm x}/\rho_{\rm x}$ is the dimensionless parameter of the equation of state related to dark energy. In the framework of IDMDE models one can assume a flat Universe with the Friedman-Lema\^itre-Robertson-Walker (FLRW) metric and pressureless dark matter, i.e., $\omega_{\rm c} = p_{\rm c}/\rho_{\rm c}=0$. Note that Eqs.~\eqref{EEMDc} and \eqref{EEMDx} can also be reformed as follows:
	\begin{equation}
	\rho \sp \prime_{\rm c} + 3 \mathcal{H} (1+\omega^{eff}_{\rm c}) = 0,
	\end{equation}
	\begin{equation}
	\rho \sp \prime_{\rm x} + 3 \mathcal{H} (1+\omega^{eff}_{\rm x}) = 0.
	\end{equation}
In these relations, $\omega^{eff}_{\rm c}$ and $\omega^{eff}_{\rm x}$ are effective parameters of dark matter and dark energy equations of state that can be obtained as follows:
	\begin{equation}
	\omega^{eff}_{\rm c} = \dfrac{a Q}{3 \mathcal{H} \rho_c},
	\end{equation}
	\begin{equation}
	\omega^{eff}_{\rm x} = \omega_{\rm x} - \dfrac{a Q}{3 \mathcal{H} \rho_{\rm x}}.
	\end{equation}
As mentioned earlier, $Q$ is the energy density transfer rate, which implies for $Q>0$ the direction of energy transfer is from dark matter to dark energy, and vice versa. In addition, the energy density transfer rate leads to the modification in the redshift evolution of dark sector components. For instance, if $Q>0$, the redshift evolution of dark matter will be faster. Thus, there could have been more dark matter at higher redshifts compared to noninteracting models \cite{LopezHonorez:2009tuj}. Actually, different forms have been proposed for the energy density transfer rate \cite{Yang:2018xlt, Yang:2018pej, Yang:2019vni, Yang:2019uzo, Yang:2019uog}. Throughout this work, we focus on $Q \propto H \rho_{\rm x}$. 
	
On the other hand, special attention should be paid to the fact that in order to provide suitable conditions for the structure formation, interacting models must be protected from instabilities that may impose to models. To this end, the stability of interacting models needs to be addressed that will be discussed in more detail in Sec.~\ref{sec:iii}.
	
\subsection{The Inhomogeneous Universe}
Although on large scales, the Universe seems simple and can be considered as an isotropic and homogeneous Universe, it contains structures such as galaxies and galaxy clusters on smaller scales. Therefore, to describe these structures, the presence of inhomogeneities seems to be necessary on smaller scale.These structures can be considered as a result of the collapse of density fluctuations that have exceeded their threshold. During the early stages of the evolutionary Universe, inhomogeneities in matter density can arise due to the gravitational instabilities. This leads to the formation of structures that can be observed in the present-time Universe. 
	
Therefore, to justify the structure formation on small scales, a perturbed metric can be considered as
\begin{equation}
	ds^2 = (\bar{g}_{\rm \mu \nu} + \delta g_{\rm \mu \nu}) dx^{\rm \mu} dx^{\rm \nu},
\end{equation}
where $\bar{g}_{\rm \mu \nu}$ demonstrates the unperturbed background metric and $|\delta g_{\rm \mu \nu}| \ll 1$ reffers to small perturbations. Given that vector perturbations decay very quickly and tensor perturbations do not induce perturbations in the perfect fluid, it makes sence to consider scalar perturbations for linear fluctuations. Hence, the perturbed FLRW metric within a homogeneous and isotropic Universe takes the following form \cite{Wang:2016lxa}
	\begin{eqnarray}
	ds^2 = a^2(\tau) \bigg[-\(1+2 \phi\) d \tau^2 + 2 \partial_{\rm i} B d \tau dx^{\rm i} +		\nonumber\\
	\[(1- 2\psi) \delta_{\rm ij} + 2 \partial_{\rm i} \partial_{\rm j} E\] dx^{\rm i} dx^{\rm j} \bigg],
	\end{eqnarray}
	where $\tau$ is the conformal time, $\phi$, $\psi$, $B$ and $E$ are the scalar metric perturbations, and $d\tau = dt / a$.
	
	Moreover, in the rest frame of fluid ``A", the sound speed, $c_{\rm s A}$, is equal to the propagation velocity of pressure fluctuations \cite{Valiviita:2008iv}:
	\begin{equation}\label{soundspeed}
	c_{\rm s A}^2 = \dfrac{\delta P_{\rm A}}{\delta \rho_A}\bigg|_{\rm rest \ frame}.
	\end{equation}
	Also, the adiabatic sound speed can be defined as \cite{Valiviita:2008iv}
	\begin{equation}\label{soundspeedadiabatic}
	c_{\rm a A}^2 \equiv \dfrac{d P_{\rm A}}{d \rho_{\rm A}} = \omega_{\rm A} + \dfrac{\omega_{\rm A} \sp{\prime}}{\rho_{\rm A} \sp{\prime}/\rho_{\rm A}}.
	\end{equation}
If a barotropic equation of state is considered, then both Eqs.~\eqref{soundspeed} and \eqref{soundspeedadiabatic} will be equal. Hence, in the presence of a constant $\omega$, the adiabatic sound speed will reduce to $c_{\rm a}^2 = \omega$. Besides, the sound speed would be imaginary if we consider dark energy as an adiabatic fluid, i.e., $c_{\rm sx}^2 = c_{\rm ax}^2 = \omega_{\rm x} < 0$, which leads to instabilities in dark energy \cite{Valiviita:2008iv}. This issue can be addressed by demanding $c_{\rm sx}^2 > 0$. It should be mentioned that in this work we set $c_{\rm sx}^2 = 1$ in the calculations.
	
	To calculate the matter power spectrum numerically within the context of interacting models, a modified version of the {\it Code for Anisotropies in the Microwave Background} (CAMB) \cite{Lewis:1999bs} must be employed. For this purpose, the code has been adopted to solve the IDMDE Friedmann equation. Note that the CAMB code solves the Friedmann equation of the $\Lambda$CDM model in its standard form. For the chosen energy density transfer rate as $Q \propto H \rho_{\rm x}$, analytical solutions for both energy densities, i.e., $\rho_{\rm c}$ and $\rho_{\rm x}$, can be specified. The CAMB code requires the energy densities as functions of the scale factor. To do this, one has to solve two coupled Eqs.~\eqref{EEMDc} and \eqref{EEMDx} as functions of scale factor. Furthermore, compared to the $\Lambda$CDM model, $Q$ modifies linear perturbations that must be considered in the modified version of the CAMB code.
	
By taking the perturbed part of Eq.~(\ref{EMT-not-conserved}) in terms of the Fourier space, evolution equations of the dimensionless density perturbation $\delta_{\rm A}=\delta\rho_{\rm A}/\rho_{\rm A}$ (the continuity equation) and of the velocity perturbation equation $\theta_{\rm A}\equiv \partial_{i}v^{i}_{\rm A}$ (the Euler equation) for derk sector components can be calculated as follows:
	\begin{eqnarray}
	\delta \sp \prime_{\rm x} = - \(1+\omega_{\rm x}\) \bigg( \theta_{\rm x} + \dfrac{h \sp \prime}{2} \bigg) - 3 \mathcal{H} \( c_{\rm sx}^2 - \omega_{\rm x}\) \nonumber \\ 
	\times \left[ \delta_{\rm x} + 3 \mathcal{H} \(1+\omega_{\rm x} \) \dfrac{\theta_{\rm x}}{k^2} \right] - 3 \mathcal{H} \omega_{\rm x} \sp \prime \dfrac{\theta_{\rm x}}{k^2}	\nonumber \\
	+ \dfrac{a Q}{\rho_{\rm x}} \left[ - \delta{\rm x} + \dfrac{\delta Q}{Q} +3 \mathcal{H} (c^2_{\rm s x} - \omega_{\rm x} ) \dfrac{\theta_{\rm x}}{k^2} \right],
	\label{del-x-p}
	\end{eqnarray}
	\begin{eqnarray}
	\theta_{\rm x} \sp \prime = - \mathcal{H} \(1-3 c^2_{\rm sx}\) \theta_{\rm x} + \dfrac{c^2_{\rm sx}}{(1+\omega_{\rm x})} k^2 \delta_{\rm x} \nonumber \\
	+ \dfrac{a Q}{\rho_{\rm x}} \left[ \dfrac{\theta_{\rm c} - \(1+c^2_{\rm sx}\) \theta_{\rm x}}{1+\omega_{\rm x}} \right],
	\label{theta-x-p}
	\end{eqnarray}
	\begin{equation}
	\delta_{\rm c} \sp \prime = - \bigg(\theta_{\rm c} + \dfrac{h \sp \prime}{2}\bigg) + \dfrac{a Q}{\rho_{\rm c}} \bigg(\delta_{\rm c} - \dfrac{\delta Q}{Q} \bigg),
	\label{del-c-p}
	\end{equation}
	\begin{equation}
	\theta_{\rm c} \sp \prime = - \mathcal{H} \theta_{\rm c}.
	\label{theta-c-p}
	\end{equation}
In these relations, $k$ is the wavenumber of Fourier mode and $h(\tau, \vec{k})$ is the synchronous field in the $k\mbox{-}$space \cite{Ma:1995ey}. As can be seen, the dark matter velocity perturbation equation is the same as the uncoupled one, since there is no momentum transfer in the dark matter frame. Thus, one can consistently set $\theta_{\rm c}=0$ in a synchronous gauge \cite{Valiviita:2008iv}.

In the next section, we will briefly discuss the stability conditions within the framework of IDMDE models.
\section{Stability and Instability}\label{sec:iii}
As mentioned earlier, the large-scale stability of IDMDE models is very important, because the Universe would not yield structure formations without stable perturbations \cite{Yang:2018xlt}. But it should be noted that IDMDE models may initially suffer from non-adiabatic instabilities on large scales. These instabilities take place due to the presence of the coupling terms in the propagation of dark energy pressure waves \cite{Gavela:2009cy, LopezHonorez:2009tuj, LopezHonorez:2009hpu}. It should be noted that Eqs.~\eqref{del-x-p}$\mbox{-}$\eqref{theta-c-p} are obtained in the rest frame of dark matter, while the sound speed of dark energy, $c^{2}_{\rm sx}$, should be evaluated in the rest frame of dark energy. As a result, one can specify the pressure perturbation in the rest frame of sound speed as follow \cite{Bean:2003fb}:
	\begin{equation}\label{pressurepert}
		\delta p_{\rm x} = c^2_{\rm s x} \delta \rho_{\rm x}-(c^2_{\rm s x} - c^2_{\rm a x})\rho_{\rm x} \sp \prime \dfrac{\theta_{\rm x}}{k^{2}}.
	\end{equation}
If there is an interaction between dark sector components (i.e., in the presence of a nonzero $Q$), the dark energy pressure perturbation in terms of the density perturbation can be deduced from Eqs.~\eqref{EEMDx} and \eqref{pressurepert} as
	\begin{eqnarray}
	\dfrac{\delta p_{\rm x}}{\delta \rho_{\rm x}} = c^2_{\rm s x} + 3 \(c^2_{\rm s x} - c^2_{\rm a x}\) \(1+\omega^{\rm eff}_{\rm x}\) \dfrac{\mathcal{H} \theta_{\rm x}}{k^2 \delta_{\rm x}} \hspace*{1.05cm}\nonumber \\
	= c^2_{\rm s x} + 3 \(c^2_{\rm s x} - c^2_{\rm a x}\) \(1+\omega_{\rm x}\)\(1+d\)\dfrac{\mathcal{H} \theta_{\rm x}}{k^2 \delta_{\rm x}},
	\end{eqnarray}
	where $d$ represents the doom factor and is defined as:
	\begin{equation}
	d \equiv - \dfrac{ a Q}{3 \mathcal{H} \rho_{\rm x} (1+\omega_{\rm x})}.
	\label{doom}
	\end{equation}
	
	In Ref.~\cite{Gavela:2009cy}, it has been stated that the stability or instability of interacting models depends on the doom factor in such a way that for $d>1$, i.e., strong coupling regime, an exponential growth of the dark energy perturbations leads to the large-scale instability. In the mentioned reference the CAMB and {\it Cosmological MonteCarlo} (CosmoMC) codes have been employed with type Ia supernova and CMB data. As a result, it has been shown that the best fit case can be obtained when $d < 0$, which is stable and consistent with observational data. In other words, the negative doom factor can guarantee the stability of IDMDE models.
	
In this work, we choose a specific functional form of the energy density transfer rate as
\begin{equation}
Q= 3 H \xi (1+\omega_{\rm x}) \rho_{\rm x},
\end{equation}
where $\xi$ is the coupling parameter. According to Ref.~\cite{Yang:2017zjs}, the novelty of the selected form for the energy density transfer rate is it includes the factor $(1+\omega_{\rm x})$. This factor can be simplified by the expression in the denominator of Eq.~\eqref{doom} and only a simple constraint on the coupling parameter remains. In other words, the model can alleviate large-scale perturbation instability if and only if the coupling parameter adheres to $\xi \ge 0$. Under these situations, once the energy density transfer rate is determined, Eqs.~\eqref{del-x-p}$\mbox{-}$\eqref{theta-c-p} can be reformed as follows:
	\begin{eqnarray}
	\delta \sp \prime_{\rm x} = - \(1+\omega_{\rm x}\) \bigg( \theta_{\rm x} + \dfrac{h \sp \prime}{2} \bigg) - 3 \mathcal{H} \( c_{\rm sx}^2 - \omega_{\rm x}\) \nonumber\\ 
	\times \bigg[ \delta_{\rm x} + 3 \mathcal{H} \(1+\omega_{\rm x} \) \dfrac{\theta_{\rm x}}{k^2} \bigg] + 3 \mathcal{H} \xi \(1+\omega_{\rm x} \) \nonumber\\
	\times\bigg[ \dfrac{\theta + h \sp \prime /2}{3 \mathcal{H}} +3 \mathcal{H} (c^2_{\rm s x} - \omega_{\rm x} ) \dfrac{\theta_{\rm x}}{k^2} \bigg],
	\end{eqnarray}
	\begin{eqnarray}
	\theta_{\rm x} \sp \prime = - \mathcal{H} \(1-3 c^2_{\rm sx}\) \theta_{\rm x} + \dfrac{c^2_{\rm sx}}{(1+\omega_{\rm x})} k^2 \delta_{\rm x} \nonumber\\
	+ 3 \mathcal{H} \xi \[\theta_{\rm c} - \( 1+c^2_{\rm s x} \) \theta_{\rm x} \],
	\end{eqnarray}
	\begin{eqnarray}
	\delta_{\rm c} \sp \prime = - \bigg(\theta_{\rm c} + \dfrac{h \sp \prime}{2}\bigg) + 3 \mathcal{H} \xi \(1+\omega_{\rm x}\) \dfrac{\rho_{\rm x}}{\rho_{\rm c}} \nonumber \\
	\times \left(\delta_{\rm c} - \delta_{\rm x} - \dfrac{\theta + \frac{h \sp \prime}{2}}{3 \mathcal{H}} \right),
	\end{eqnarray}
	\begin{equation}
	\theta_{\rm c} \sp \prime = - \mathcal{H} \theta_{\rm c}.
	\end{equation}

Up to here, we have described the theoretical framework and stability conditions of IDMDE models. In the following, we will calculate the ISW effect within the context of such models.

\section{The Integrated Sachs-Wolfe Effect}\label{sec:iv}
From the last scattering surface to us, CMB photons move through the gravitational potential of the matter, like the galaxy clusters. When photons move into a gravitational potential well, they become blueshifted. On the other hand, they will be redshifted as they move out of the gravitational potential. During about the $13.8$ billion years it takes the first free photons to reach us, they pass through different gravitational potential wells. Therefore, the shifts we see will be accumulated along the observer's line of sight, which can potentially reveal some phenomenological aspects of the late-time Universe. 

To derive the ISW power spectrum, one can start with the matter power spectrum. To do this, we modify the publicly available CAMB package \cite{Lewis:1999bs}, and employ mean values of Table~\ref{Mean-value}. The second column of this table shows the cosmological parameters of the $\Lambda$CDM model, derived from the CMB temperature fluctuations measured by Planck with the addition of external data sets. Error bars are given at $68\%$ confidence level \cite{Planck:2015fie}, and the third column reveals the mean values of the free and derived cosmological parameters with their errors at $68.3\%$ and $95.4\%$ confidence regions for the IDMDE model \cite{Yang:2017zjs}.
	\begin{table}
	\begin{center}
		\caption{Mean values of the free cosmological parameters for the $\Lambda$CDM and IDMDE models.}
		\label{Mean-value}
		\begin{tabular}{c|c|c}
				\hline
					\hline
			Parameter & $\Lambda$CDM model & IDMDE model\\
			\hline
			$\Omega_{\rm b} h^2$ & $0.02230 \pm 0.00014$ &  $0.0223^{+0.0001+0.0003}_{-0.0001-0.0003}$\\
			$\Omega_{\rm c} h^2$ & $0.1188 \pm 0.0010$ & $0.1183^{+0.0014+0.0030}_{-0.0014-0.0029}$ \\
			$100\,\theta_{\rm MC}$ & $1.04093 \pm 0.00030$ & $1.0406^{+0.0003+0.0006}_{-0.0003-0.0006}$ \\
			$\tau$ & $0.066 \pm 0.012$ & $0.0663^{+0.0161+0.0315}_{-0.0162-0.0319}$ \\
			$n_{\rm s}$ & $0.9667 \pm 0.0040$ & $0.9760^{+0.0036+0.0071}_{-0.0038-0.0070}$ \\
			$\ln(10^{10}A_{\rm s})$  & $3.064 \pm 0.023$ & $3.0722^{+0.0311+0.0605}_{-0.0288-0.0616}$ \\
			$H_{\rm 0}$ & $67.74 \pm 0.46$ & $68.4646^{+0.8199+1.3348}_{-0.7380-1.3616}$ \\			
			$\omega_{\rm x}$ & $-1.000$ &  $-1.0230^{+0.0329+0.0527}_{-0.0257-0.0603}$ \\			
			$\xi$ & - & $0.0360^{+0.0091+0.0507}_{-0.0360-0.0360}$ \\
				\hline
					\hline
		\end{tabular}
	\end{center}
\end{table}

\begin{figure}
	\includegraphics[width=\columnwidth]{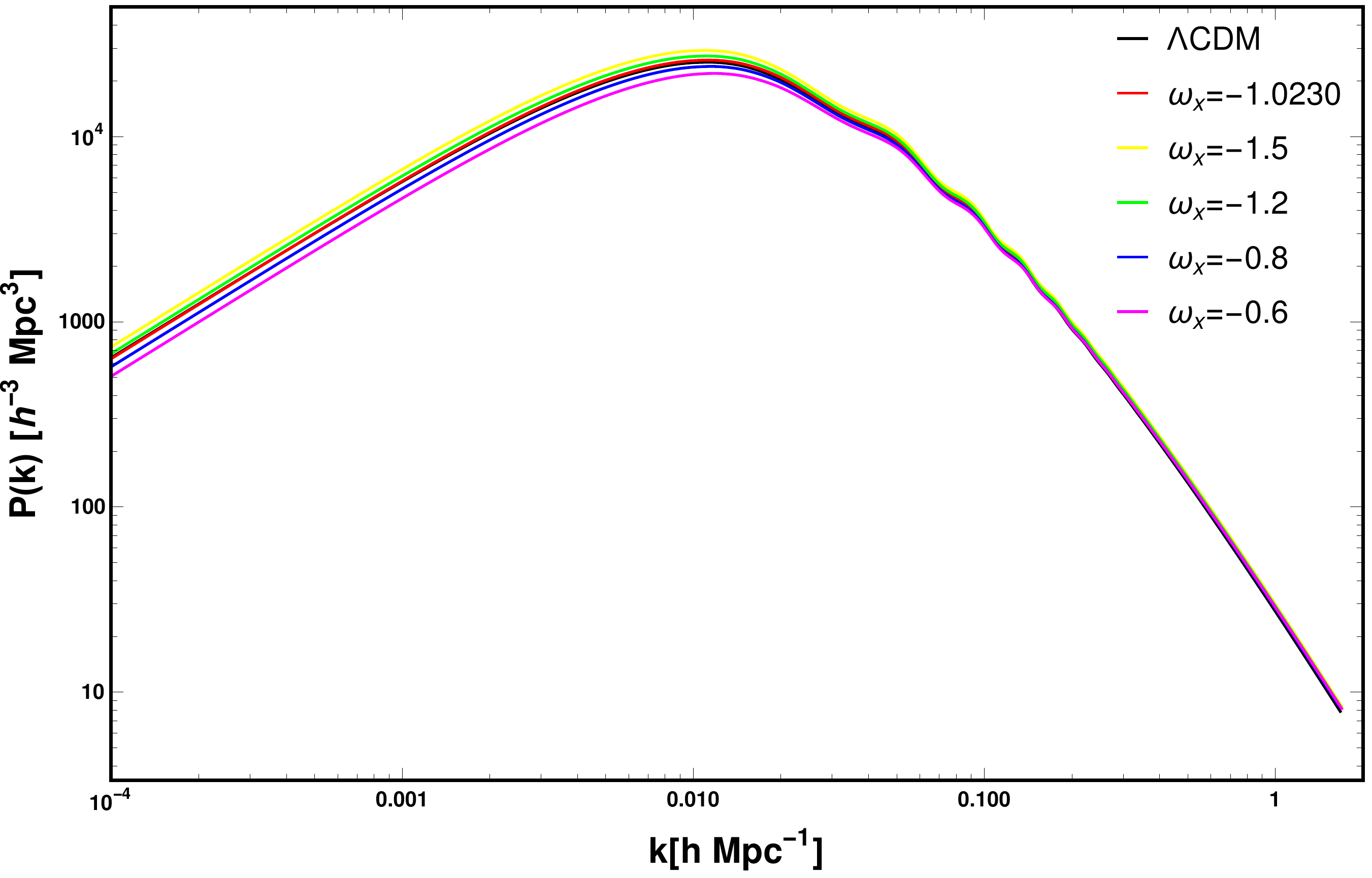}
\\~\\~\\
	\includegraphics[width=\columnwidth]{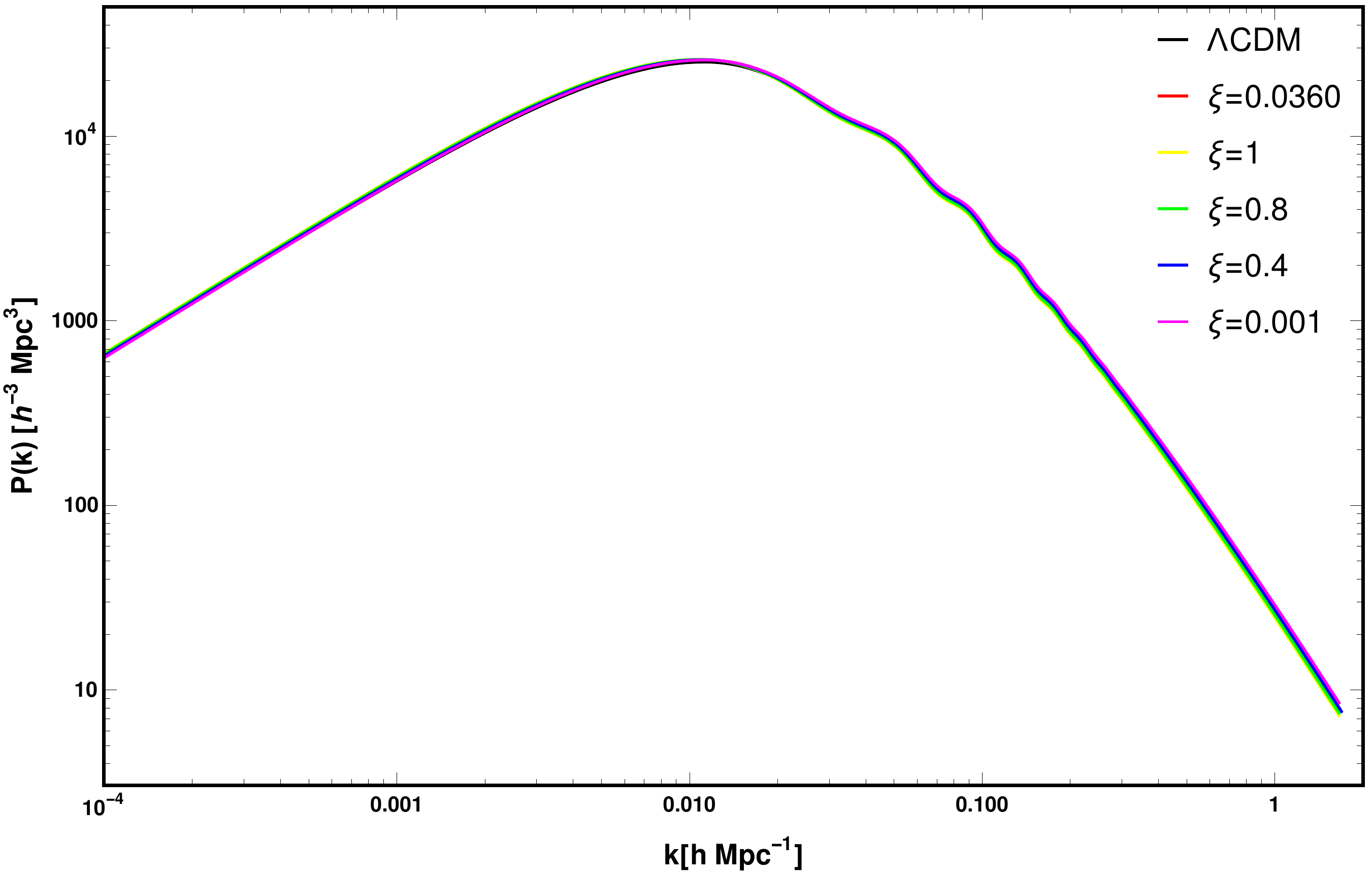}
	\caption{The matter power spectrum as a function of comoving wavenumber $k$ for the IDMDE and $\Lambda$CDM models.{{\bf{Top:}} For different values of the dimensionless parameter of dark energy equation of state $\omega_{\rm x}$}. {\bf{Bottom:}} For different values of the coupling parameter $\xi$. For the upper panel, the coupling paramete is kept fixed as $\xi=0.0360$, and for the lower panel, the dimensionless parameter of dark energy equation of state is set to be $\omega_{\rm x}=-1.0230$.}
	\label{MPS}
\end{figure}

In Fig.~\ref{MPS}, we have shown the matter power spectrum in the present-time Universe, while considering the IDMDE model, and have compared it with the corresponding result obtained from $\Lambda$CDM model. The top panel of this figure is plotted for different values of the dimensionless parameter of the dark energy equation of state. As can be seen from this figure, the peak position of the matter power spectrum is shifted toward larger scales (i.e., toward smaller $k$) in $\omega_{\rm x} = -1.5$ and increasing in the matter power spectrum. Also, the peak of the matter power spectrum is shifted toward smaller scales (i.e., toward larger $k$) in $\omega_{\rm x} = -0.6$ and decreasing in the matter power spectrum. In addition, in the bottom panel of this figure, similar calculations are performed for different values of the coupling parameter $\xi$. As it turns out, the coupling parameter $\xi$ does not yield meaningful changes in the matter power spectrum of the IDMDE model, and only a very slight deviation from the $\Lambda$CDM model is distinguishable on small scales.

Moreover, it is known that the gravitational potential is constant for a matter-dominated Universe and therefore does not yield ISW signal. Hence, just as we are looking for it to be, the ISW effect can serve as an indicator of something different from ordinary pressureless matter. Moreover, the late-time ISW effect is referred to the integrated differential gravitational redshift caused by the evolution of gravitational potentials along the path traveled by photons. In this regard, on large angular scales, the ISW effect produces a weak positive correlation for all frequencies \cite{SDSS:2003lnz}.

On the other hand, the ISW effect is the main secondary anisotropy at large angles (i.e., at low $l$) that reflects the decay of the gravitational potential. A direct way to specify the late-time ISW effect is to cross-correlate the large-angle CMB anisotropies with large-scale structures at low redshifts such as angular galaxy correlations. Therefore, the temperature anisotropy due to the ISW effect is specified by taking an integration on the time-varing quantity, ($\phi - \psi$), along the line-of-sight \cite{Schaefer:2008qs, Y:2021ybx}, i.e.,

\begin{eqnarray}\label{t-isw}
\Theta_{\rm ISW} = \dfrac{\Delta T}{T_{\rm CMB}}\bigg|_{\rm ISW} = \dfrac{2}{c^2} \int^1_{\rm a_{\rm dec}} \dfrac{\partial \phi}{\partial a} da  \hspace*{1.6cm}\nonumber \\
= - \dfrac{2}{c^3} \int_0^{\rm \chi_{\rm H}} a^2 H(a) \dfrac{\partial \phi}{\partial a} d \chi,
\end{eqnarray}
where $T_{\rm CMB} = 2.725 \ K$ is the CMB temperature, $a_{\rm dec}$ is the time at which photons decouple, $c$ is the velocity of light in vacuum. Also, $\chi$ is the comoving distance between the CMB and us that can be calculated as a function of scale factor, namely
\begin{equation}
\chi(a) = \int_0^1 \dfrac{c \ da}{a^2 H(a)}.
\end{equation}

Moreover, if the perturbations are sufficiently within the horizon, the Poisson's equation would allow us to relate the matter fluctuation field, $\delta_{\rm m}$, to the potential at late times \cite{Manzotti:2014kta, Stolzner:2017ged}. Hence, one can have 
\begin{equation}
\phi (k,a)=  - \dfrac{3}{2} \ a^2 \ H^2(a) \ \Omega_{\rm m} (a) \ \dfrac{\delta_{\rm m}(k,a)}{k^2},
\label{poisson}
\end{equation}
where $\Omega_{\rm m} (a)$ is the dimensionless matter density parameter. In this regard, by using Eqs.~\eqref{t-isw} and \eqref{poisson}, the temperature anisotropy due to the ISW effect can take the following form:
\begin{equation}
\Theta_{\rm ISW} = \dfrac{3}{c^3} \int^{\rm \chi_{\rm H}}_{0} a^2 H(a) \dfrac{\partial \mathcal{F}(a)}{\partial a} \dfrac{\delta_{\rm m}(k,a=1)}{k^2} d \chi.
\label{t-isw-chi}
\end{equation}
In this relation, $\mathcal{F}(a)$ is defined as
\begin{equation}\label{ffracscale}
\mathcal{F}(a) \equiv a^2 H^2(a) \Omega_{\rm m}(a) D_{\rm +}(a),
\end{equation}
where $D_{\rm +}(a) = \delta_{\rm m}(k,a)/\delta_{\rm m}(k,a=1)$ is the linear growth factor.

\begin{figure}
	\includegraphics[width=\columnwidth]{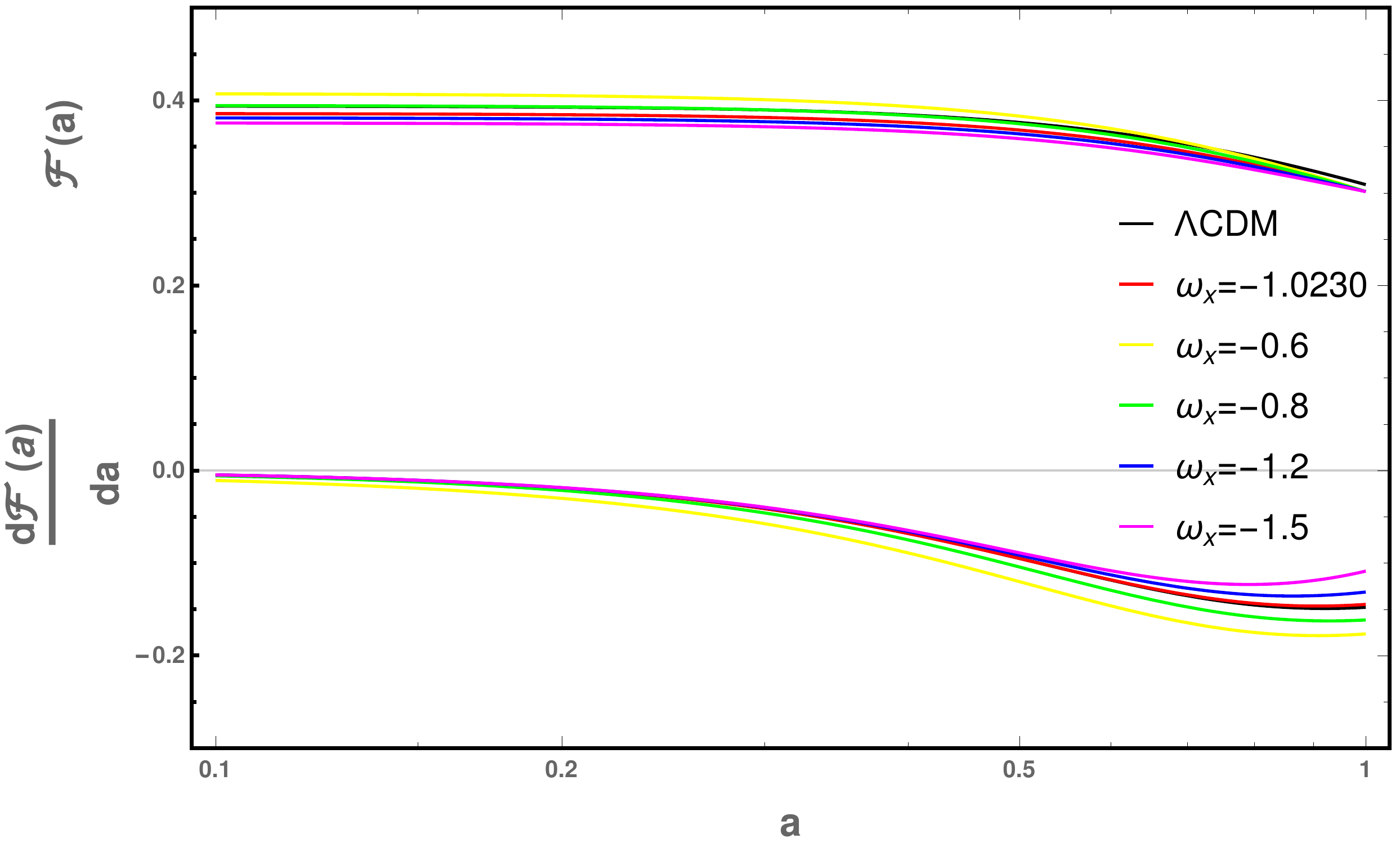}
\\~\\~\\
	\includegraphics[width=\columnwidth]{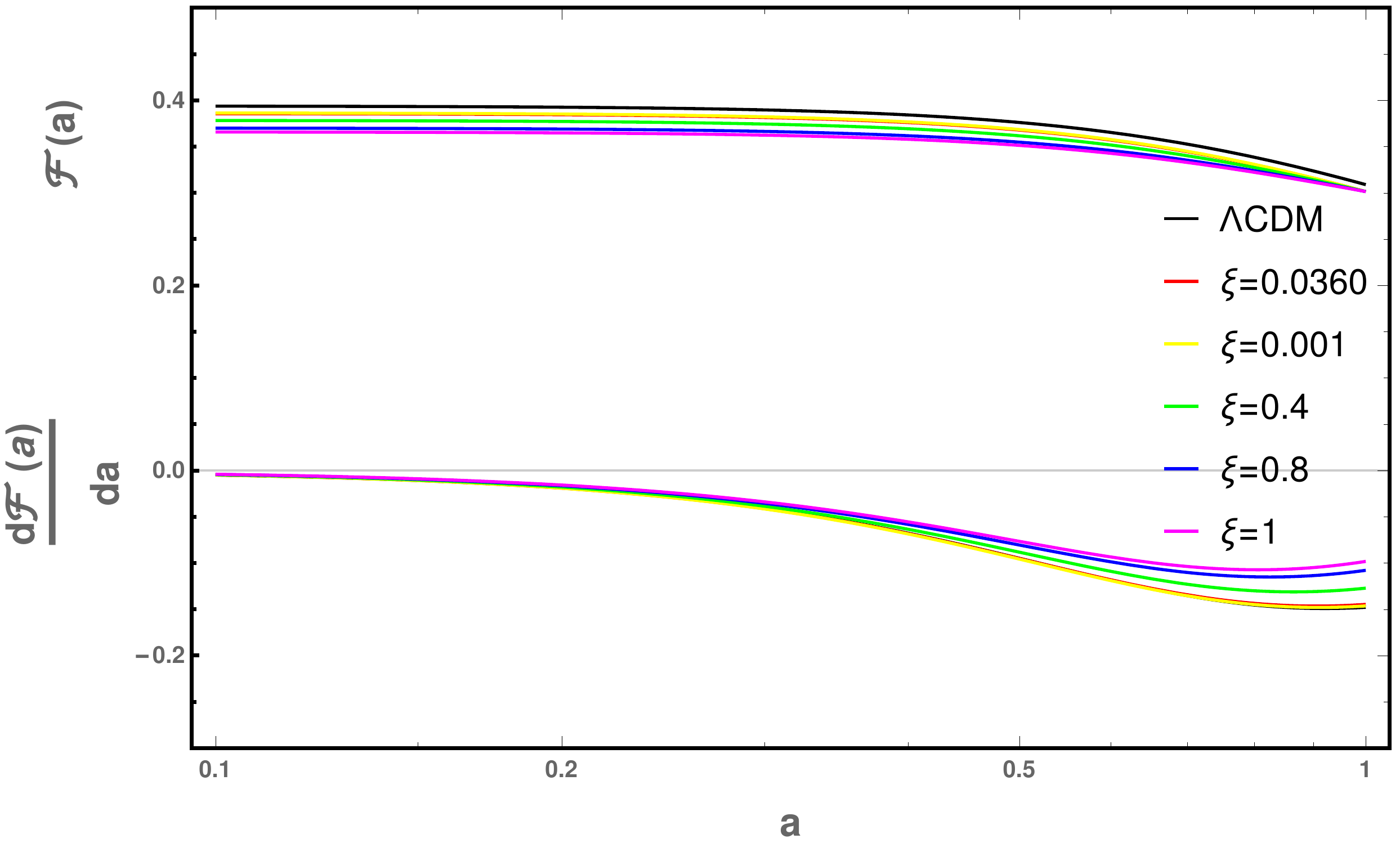}
	\caption{The time evolution $\mathcal{F}(a)$ of the ISW source term and its derivative $d\mathcal{F}(a)/da$ as functions of scale factor $a$, for the IDMDE and $\Lambda$CDM models. {{\bf{Top:}} For different values of the dark energy equation of state $\omega_{\rm x}$}. {\bf{Bottom:}} For different values of the coupling parameter $\xi$.}
	\label{F-fig}
\end{figure}

In Fig.~\ref{F-fig}, we have depicted the time evolution of the ISW source term, $\mathcal{F}(a)$, and its derivative with respect to the scale factor, $d\mathcal{F}(a)/da$, as functions of scale factor. It is clear from Eq.~\eqref{ffracscale} that in the present-time Universe (i.e., at $a=1$)‌, both models converge at $\mathcal{F}(a)=\Omega_{\rm m,0}$. 

It can be turned out form the top panel of this figure that for the case of phantom dark energy (i.e., for $\omega_{\rm x}<-1$), compared to the result obtained for the $\Lambda$CDM model, lower values of $\mathcal{F}(a)$ can be achieved at higher redshifts within the framework of the IDMDE model. Also, the quintessence dark energy (i.e., $\omega_{\rm x}>-1$) in the IDMDE model leads to higher values of $\mathcal{F}(a)$ than that obtained from the $\Lambda$CDM one. Additionally, $d\mathcal{F}/da$ in the IDMDE model is similar to the corresponding result obtained from the $\Lambda$CDM model at high redshifts , while it deviates from the $\Lambda$CDM model at low redshifts. On the other hand, the bottom panel of this figure demonstrates that at higher redshifts for all nonvanishing values of the coupling parameter, $\mathcal{F}(a)$ in the IDMDE model takes lower values than the corresponding result obtained from the $\Lambda$CDM model.

If the dark sectors are not coupled, matter density will be proportional to $\rho_{\rm m}(a) \propto a^{-3}$. Hence, one can have the following relation for the matter density
\begin{equation}
\dfrac{\Omega_{\rm m}(a)}{\Omega_{\rm m,0}} = \dfrac{H_0^2}{a^3 H^2(a)}.
\end{equation}
However, in the framework of IDMDE models the above equation cannot be used. Therefore, one has to solve the modified conservation equations for the IDMDE model (i.e., Eq.~\eqref{mod-cons-eq}). Due to this consideration, the matter density for the IDMDE model can be represented as follows:
\begin{equation}
\dfrac{\Omega_{\rm m}(a)}{\Omega_{\rm m}} =
\dfrac{H_0^2}{H^2(a)} 
\dfrac{\bigg[\rho_{\rm c,0} + \rho_{\rm b,0} - \dfrac{-1+a^{3(\omega(-1+\xi)+\xi)}}{\omega(-1+\xi)+\xi} \rho_{\rm x,0} \bigg]}{ \big(\rho_{\rm c,0} + \rho_{\rm b,0}\big) a^3}.
\end{equation}

It is known that on small scales, the ISW effect should be small. Nevertheless, the most significant ISW effects usually occur in large-scale potentials, where cosmic variance is most problematic. In other words, the ISW effect is severely limited when it comes to providing information about dark energy due to cosmic variance at large angular scales. Although it is not ideal, galaxies can be the convenient option for tracer of large-scale structures in the late-time Universe. An analysis of the correlation between ISW temperature and galaxy density can reveal potential wells and hills that may lead to anisotropies \cite{boughn2004}. In addition, in order to better understand galaxy clusters, large-scale structure surveys should be employed that calculate the abundance of galaxy clusters as a function of redshift. Correspondingly, the density of galaxies can be calculated using the following integral
\begin{eqnarray}
\delta_{\rm g} = \int \dfrac{H}{c} f(z) \delta_{\rm m}(k,a) d \chi,
\label{del-g}
\end{eqnarray}
where $f(z) = b(z) dN/dz$ is the redshift distribution function of the observed samples, $b(z)$ is the galaxy bias that relates the visible matter distribution to dark matter, and $dN/dz$ is the redshift distribution function of the survey in the comoving distance. Moreover, by using of a CMB map and a survey of galaxies, one can define the angular auto-correlation and cross-correlation as
\begin{eqnarray}
C^{\rm TT} \equiv \big< \Theta_{\rm ISW} \ \Theta_{\rm ISW} \big>,	\nonumber \\
C^{\rm Tg} \equiv \big< \Theta_{\rm ISW} \ \delta_{\rm g} \big>. 
\label{c-m-n}
\end{eqnarray}
Note that in the absence of massive neutrinos, the evolution of density fluctuations is separable in wavenumber $k$ and redshift $z$ on linear scales. Hence, by employing Eqs.~\eqref{t-isw-chi}, \eqref{del-g}, and \eqref{c-m-n}, angular auto-correlation and cross-correlation can be obtained as follows:
\begin{equation}
\begin{array}{c} \vspace{0.5cm}
C^{\rm TT}_{\rm ISW} (\mathit{l}) = \bigintss_0^{\rm \chi_{\rm H}} \dfrac{W_{\rm T}^2(\chi)}{\chi^2} \dfrac{H_0^4}{k^4} P(k = \dfrac{\mathit{l} + 1/2}{\chi}) d \chi,   \\
C^{\rm Tg} (\mathit{l}) = \bigintss_0^{\rm \chi_{\rm H}} \dfrac{W_{\rm T}(\chi) W_{\rm g}(\chi)}{\chi^2} \dfrac{H_0^2}{k^2} P(k = \dfrac{\mathit{l} + 1/2}{\chi}) d \chi,
\end{array}
\end{equation}
where $P(k)$ is the present-time matter power spectrum, $W_{\rm \mathit{l}}$ is the weight function, and $k = (l + 1/2)/\chi$ is obtained through the Limber approximation of the small angle (i.e., large $\mathit{l}$) \cite{Stolzner:2017ged, Afshordi2003-ku}.
\subsection{Redshift Distribution}
The redshift distribution of galaxies in galaxy surveys is crucial for cosmological analyses. In this regard, a spectroscopic and/or multi-band photometric calibration survey of a small patch of sky typically yields this information \cite{Sanchez2020-tj}. As mentioned earlier, to compute the cross-correlation between CMB and large-scale structure, the redshift distribution function of the observed galaxies seems to be necessary. It is important to note that each survey has a unique photometric redshift distribution and histogram function. In this work, four surveys are considered:

\begin{figure}[t!]
	\includegraphics[width=\columnwidth]{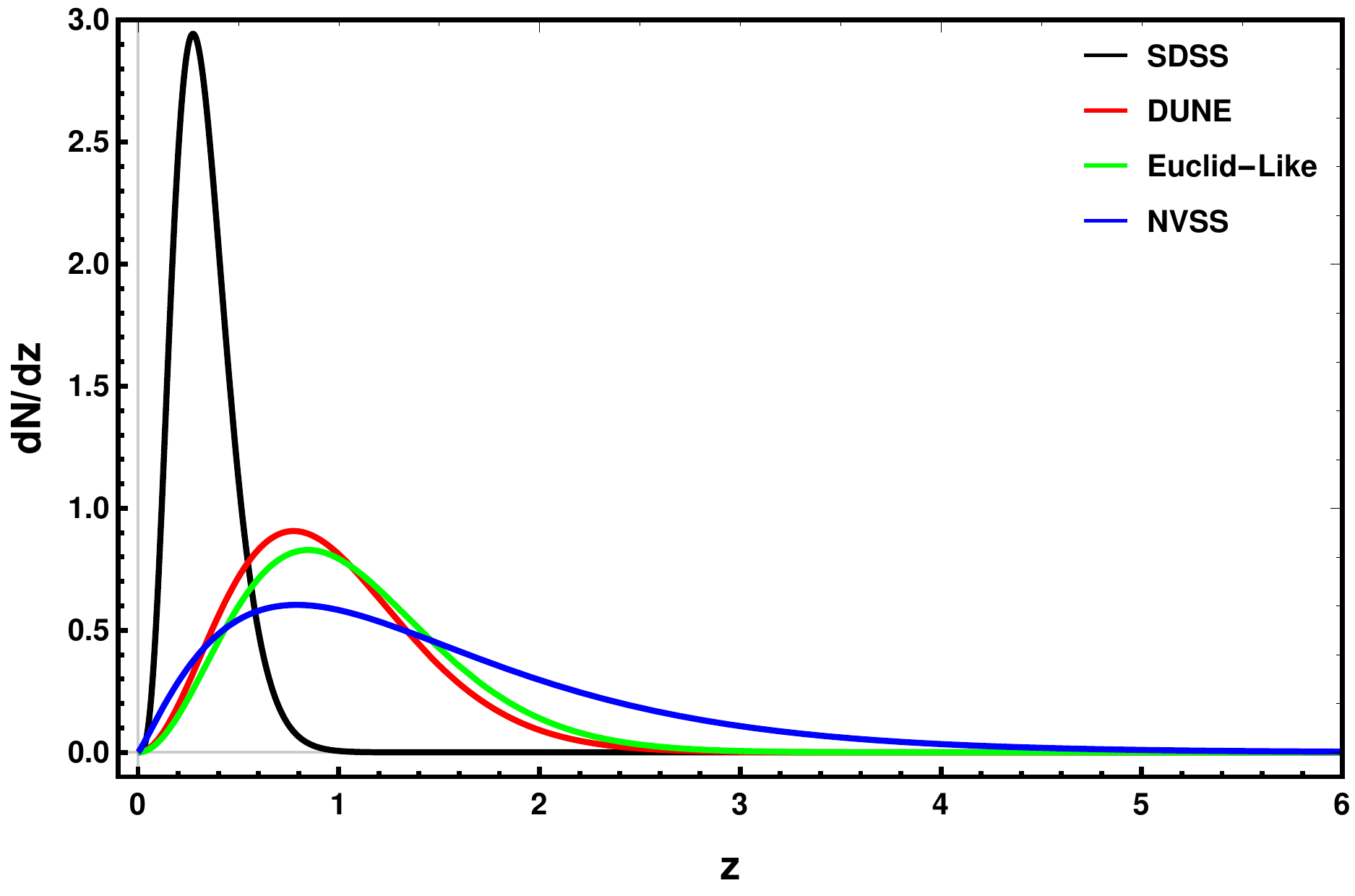}
	\caption{Photometric redshift distributions for the four catalogs used for the cross-correlation ($dN/dz$ curves are normalised to unity).}
\label{red-dist}
\end{figure}

({\bf i}) Dark Universe Explorer (DUNE) survey is designed to detect both visible light and three near-infrared bands. The main mission of this survey is to probe weak gravitational lensing candidates and the ISW effect. The redshift distribution function of this survey is defined as follow:
\begin{equation}\label{fdune}
f_{\rm DUNE} (z)  = b_{\rm eff} \left[ \dfrac{z_*}{\alpha_*} \Gamma(\dfrac{3}{\alpha_*}) \right]^{-1} \bigg( \dfrac{z}{z_*}  \bigg)^{2} \exp \left[ - \left(  \dfrac{z}{z_*} \right)^{\alpha_*} \right],
\end{equation}
where $b_{\rm eff}$, $z_{*}$ and $\alpha_{*}$ are free parameters to be determined, and $\Gamma(x)$ is the Gamma funcion.

({\bf ii}) The National Radio Astronomy Observatory (NRAO) Very Large Array (VLA) Sky Survey (NVSS) is a continuum survey that is covering about $82\%$ of the sky. The redshift distribution function of this survey has the following form:
\begin{equation}
f_{\rm NVSS} (z)  = b_{\rm eff}\, \dfrac{\alpha_*^{\alpha_* + 1}}{\Gamma (\alpha_*)} \frac{z^{\alpha_*}}{z_*^{\alpha_* +1}}\exp \left(-\dfrac{\alpha_* z}{z_*} \right).
\end{equation}

({\bf iii}) The Sloan Digital Sky Survey (SDSS) is an imaging and spectroscopic survey that collect galaxies, quasars, and stars. The redshift distribution of the main photometric SDSS galaxy sample (SDSS-MphG) is specified as fllow:
\begin{equation}
f_{\rm SDSS} (z)  = b_{\rm eff}\, \dfrac{\alpha_*}{\Gamma (\frac{m+1}{\alpha_*})} \frac{z^m}{z_*^{m+1}} \exp \left[ - \left(\dfrac{z}{z_*}\right)^{\alpha_*} \right].
\end{equation}

({\bf iv}) The Euclid-like mission will investigate the distance-redshift relations of galaxies out to redshift $z\sim 2$. It is currently scheduled to launch in 2023. For this survey, the redshift distribution takes the following form~\cite{Weaverdyck2017-nw, Martinet:2015wza}
\begin{equation}\label{feuclidlike}
f_{\rm Euclid-like} (z)  = b_{\rm eff}\ \dfrac{3}{2z_{\rm *}^3} z^2 \exp \left[ - \left(\dfrac{z}{z_{\rm *}} \right)^{3/2} \right].
\end{equation}

\begin{table}
	\begin{center}
		\caption{Redshift distribution parameters for DUNE, NVSS, SDSS and Euclid-like surveys.}
		\label{survey-params}
		\begin{tabular}{c|c|c|c|c}
			\hline
			\hline
			\hspace*{0.25cm} Survey \hspace*{0.25cm}&\hspace*{0.25cm} $b_{\rm eff}$ \hspace*{0.25cm}&\hspace*{0.25cm} $z_{*}$ \hspace*{0.25cm}&\hspace*{0.25cm} $\alpha_{*}$ \hspace*{0.25cm}&\hspace*{0.25cm} $m$ \hspace*{0.25cm}\\
			\hline
			DUNE & $1.00$ & $0.640$ & $1.500$ & - \\
			NVSS & $1.98$ & $0.790$ & $1.180$ & - \\
			SDSS & $1.00$ & $0.113$ & $1.197$ & $3.457$ \\
			Euclid-like & $1.00$ & $0.700$ & - & - \\
			\hline
			\hline
		\end{tabular}
	\end{center}
\end{table}
 
As can be seen from Eqs.~\eqref{fdune}-\eqref{feuclidlike}, there are some free parameters that should be specified. To this end, the best-fit values of these parameters have been presented in Table~\ref{survey-params}.

In Fig.~\ref{red-dist}, we have also shown the photometric normalized redshift distribution for the mentioned surveys as a function of redshift. Obviously, the NVSS survey yields the widest redshift coverage, whereas the SDSS survey leads to the narrowest redshift coverage. Also, DUNE and Euclid-like surveys provide an intermediate range of redshifts coverage. For more information such as sky coverage, number of sources, and the analytical approximation, see \cite{Stolzner:2017ged, Martinet:2015wza, Weaverdyck2017-nw, Manzotti:2014kta, Condon-NRAO, Wang:2010zzj, Douspis:2008xv, Schiavon:2012fc}.

\subsection{The ISW Power Spectrum}\label{sec:res}
\begin{figure}[t!]
\includegraphics[width=\columnwidth]{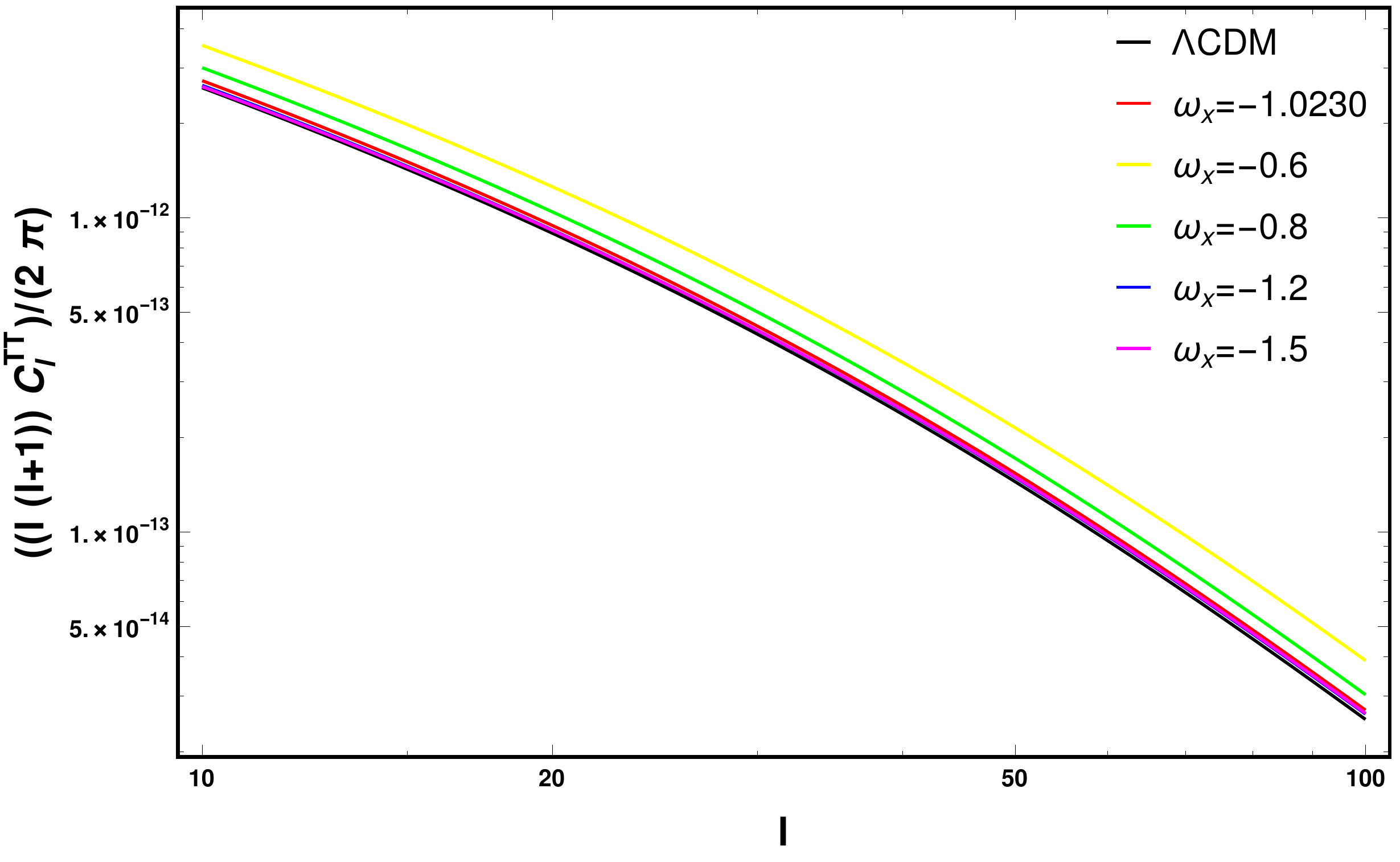}
\\~\\
\includegraphics[width=\columnwidth]{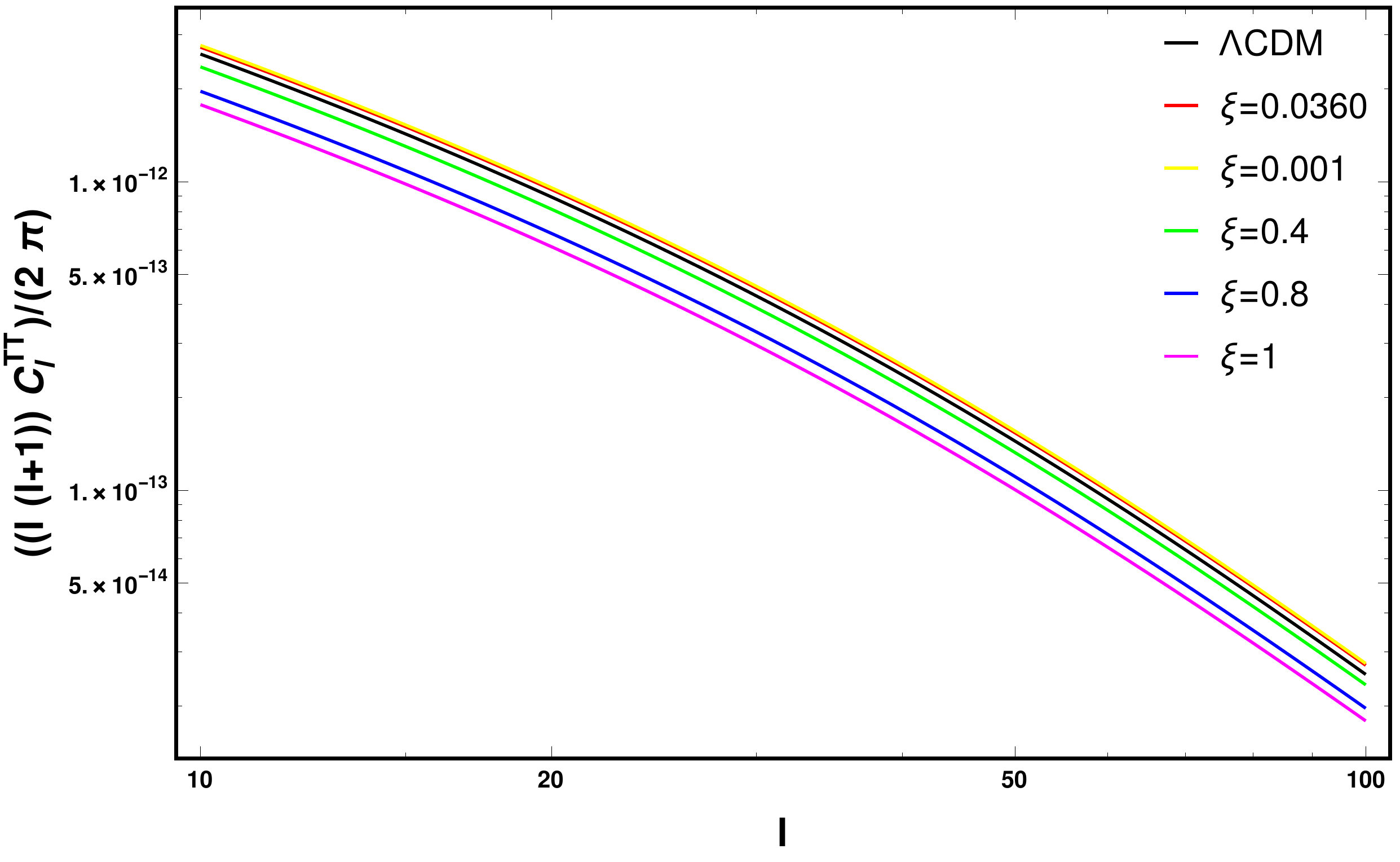}
\caption{The amplitude of the ISW-auto power spectrum as a function of multipole order $l$ for the IDMDE and $\Lambda$CDM models. {{\bf{Top:}} For different values of the dark energy equation of state $\omega_{\rm x}$}. {\bf{Bottom:}} For different values of the coupling parameter $\xi$.}
\label{CTT-w}
\end{figure}

In Fig.~\ref{CTT-w}, we have shown the amplitude of the ISW-auto power spectrum $C_{\rm TT}^{\rm ISW}$ as a function of multipole order $l=(10-100)$ for the IDMDE and $\Lambda$CDM models. The black lines show the theoretical prediction for the $\Lambda$CDM model, while the red lines indicate the theoretical prediction for the IDMDE model with the best fit values. As can be seen from the figure, the IDMDE model behaves similar to the $\Lambda$CDM model for different phantom dark energy equations of state, whereas, for the quintessence dark energy equations of state, the amplitude of the ISW-auto power spectrum for the IDMDE model is higher than that obtained from the $\Lambda$CDM model. Additionally, we have demonstrated the results for different values of the coupling parameter $\xi$. The results indicate that the value of the coupling parameter $\xi$ is inversely proportional to the amplitude of the ISW-auto power spectrum in the IDMDE model.

\begin{figure}[t!]
\includegraphics[width=\columnwidth]{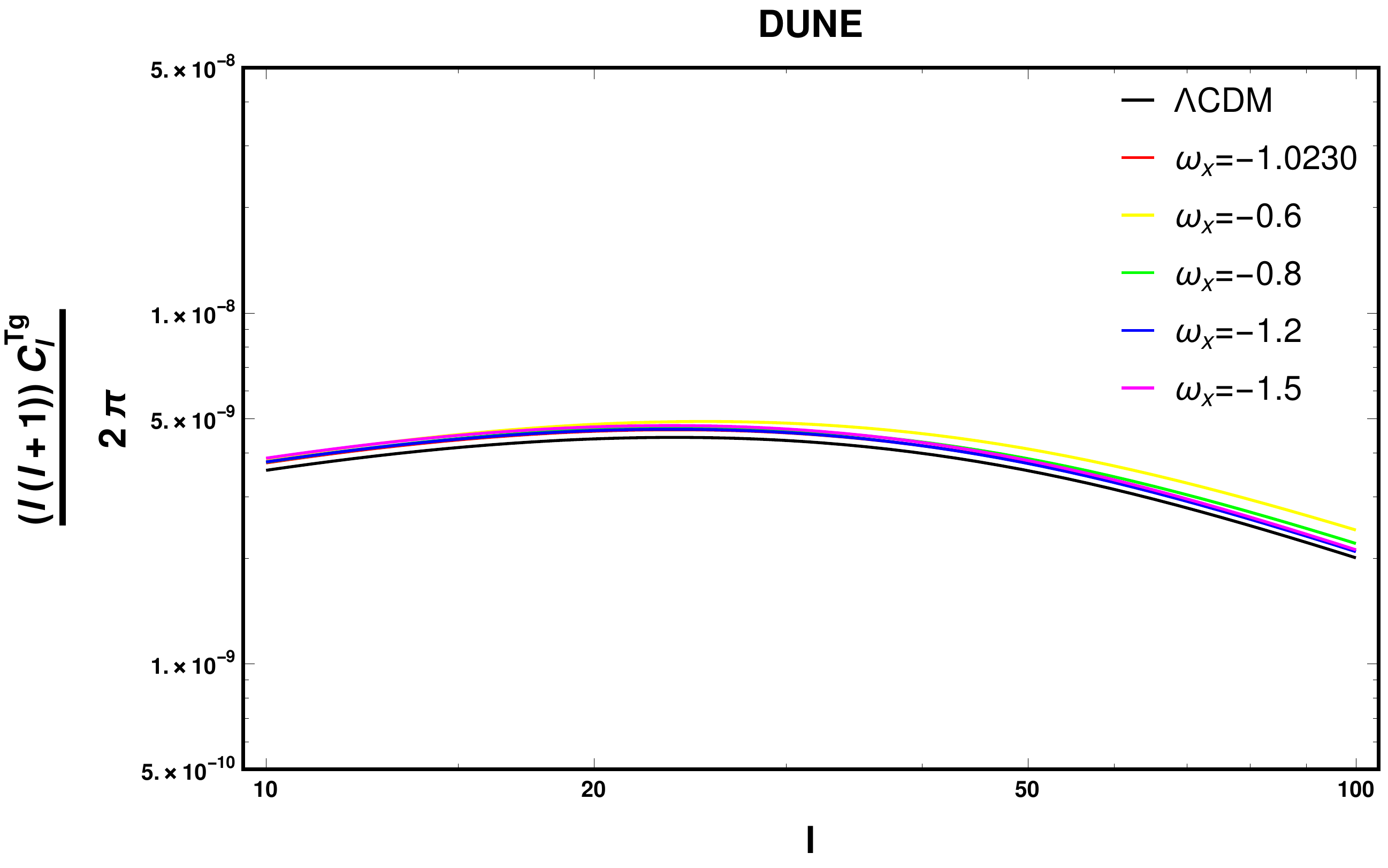}
\\~\\
\includegraphics[width=\columnwidth]{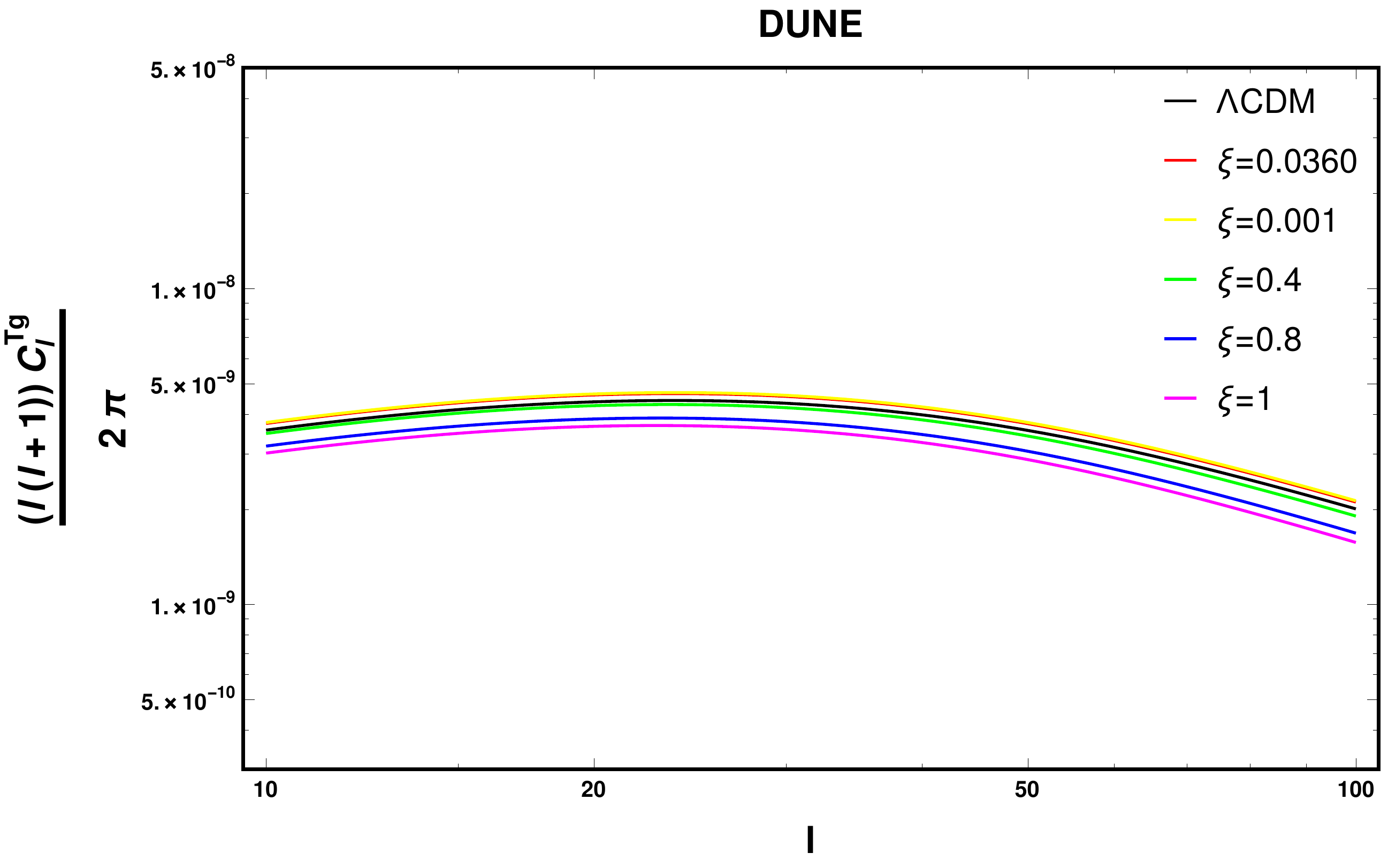}
\caption{The amplitude of the ISW cross power spectrum as a function of multipole order $l$ for the IDMDE and $\Lambda$CDM models. The DUNE survey has been considered. {{\bf{Top:}} For different values of the dark energy equation of state $\omega_{\rm x}$}. {\bf{Bottom:}} For different values of the coupling parameter $\xi$.}
\label{CTg-DUNE-w}
\end{figure}

Moreover, in Fig.~\ref{CTg-DUNE-w}, we have depicted the amplitude of the ISW-cross power spectrum as a function of multipole order $l=(10-100)$ for the DUNE survey, while considering the IDMDE model, and have compared it with the corresponding result obtained from the $\Lambda$CDM model. As can be seen from the figure, the amplitude of the ISW-cross power spectrum for the IDMDE model for all values of $\omega_{\rm x}$ is higher than that obtained from the $\Lambda$CDM model. The results exhibit that for the case of quintessence dark energy, e.g., for $\omega_{\rm x}=-0.6$, the amplitude of the ISW-cross power spectrum in the IDMDE model can be in the maximum state. Also, it can be inferred from the figure that the amplitude of the ISW-cross power spectrum in the IDMDE model changes inversely with the value of coupling parameter $\xi$.

\begin{figure}[t!]
\includegraphics[width=\columnwidth]{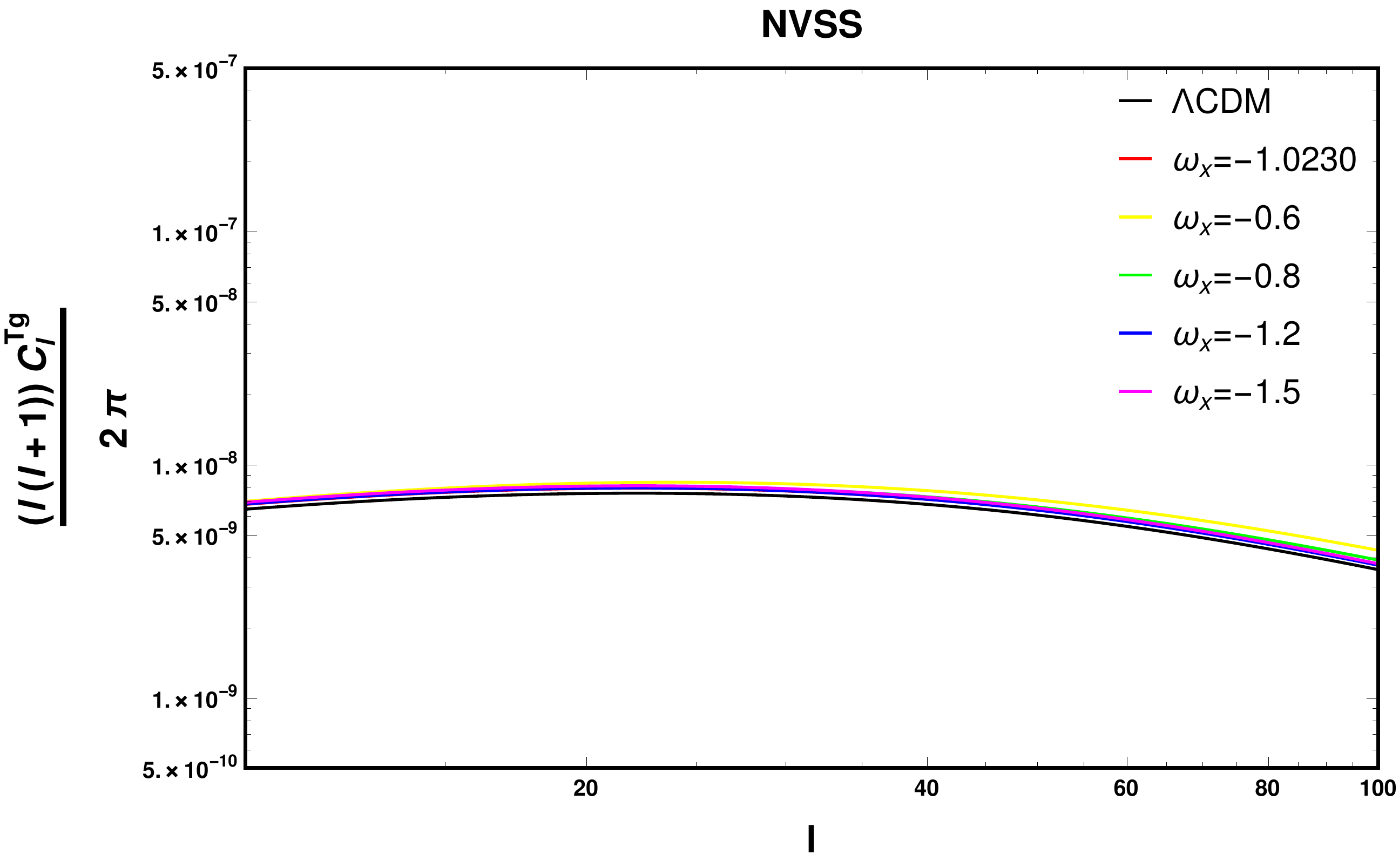}
\\~\\
\includegraphics[width=\columnwidth]{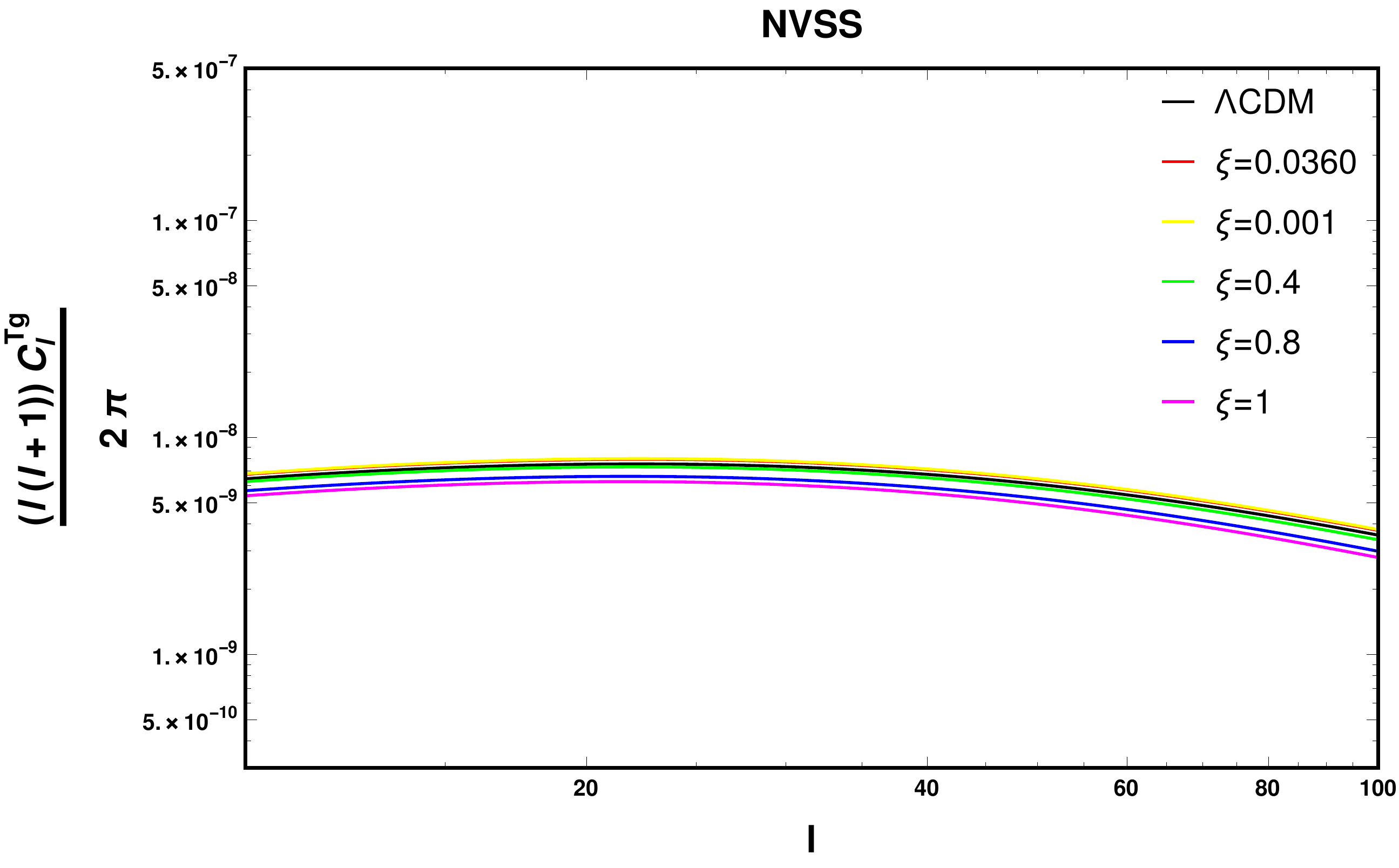}
\caption{The amplitude of the ISW cross power spectrum as a function of multipole order $l$ for the IDMDE and $\Lambda$CDM models. The NVSS survey has been considered. {{\bf{Top:}} For different values of the dark energy equation of state $\omega_{\rm x}$}. {\bf{Bottom:}} For different values of the coupling parameter $\xi$.}
	\label{CTg-NVSS-w}
\end{figure}

\begin{figure}[t!]
\includegraphics[width=\columnwidth]{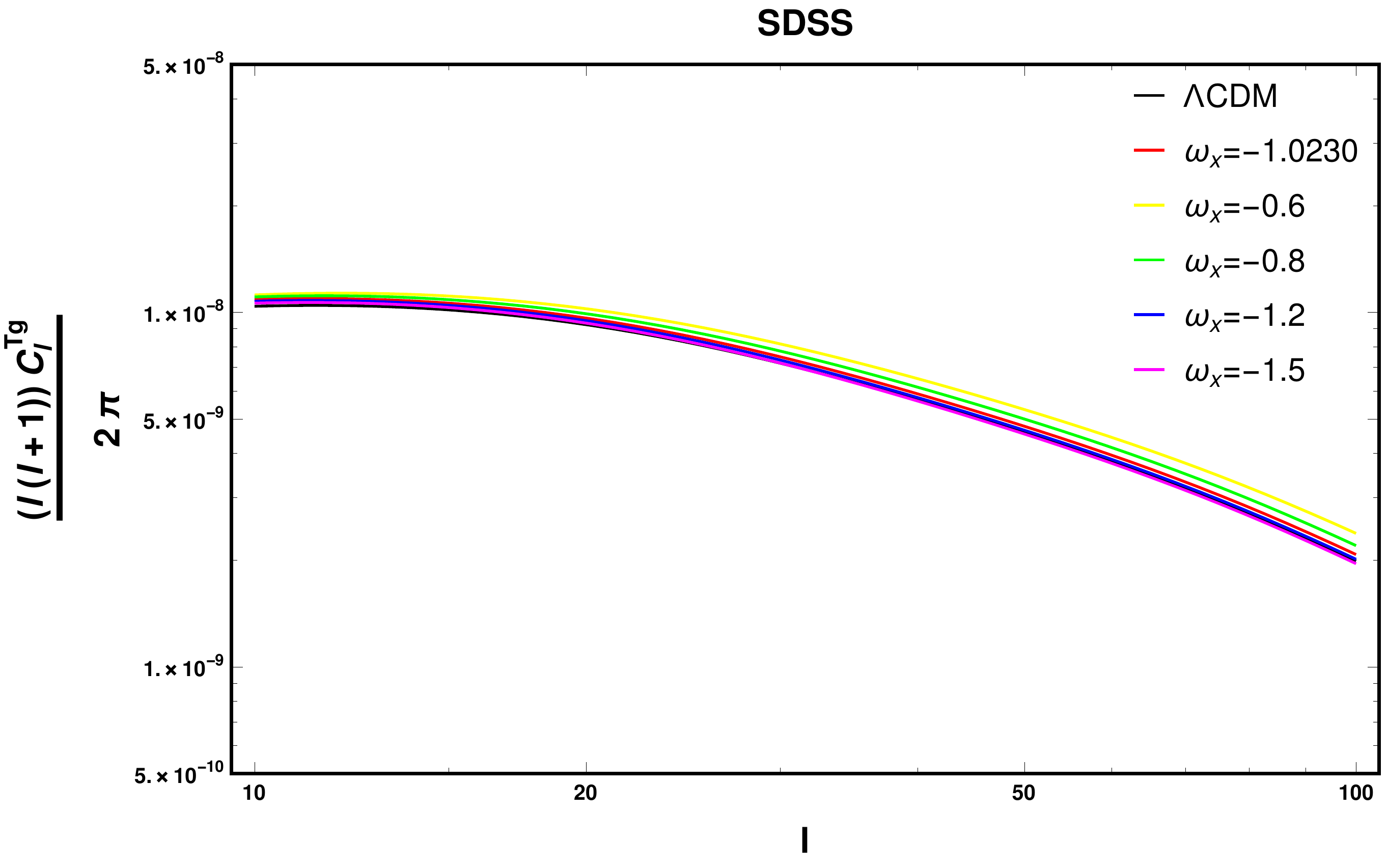}
\\~\\
\includegraphics[width=\columnwidth]{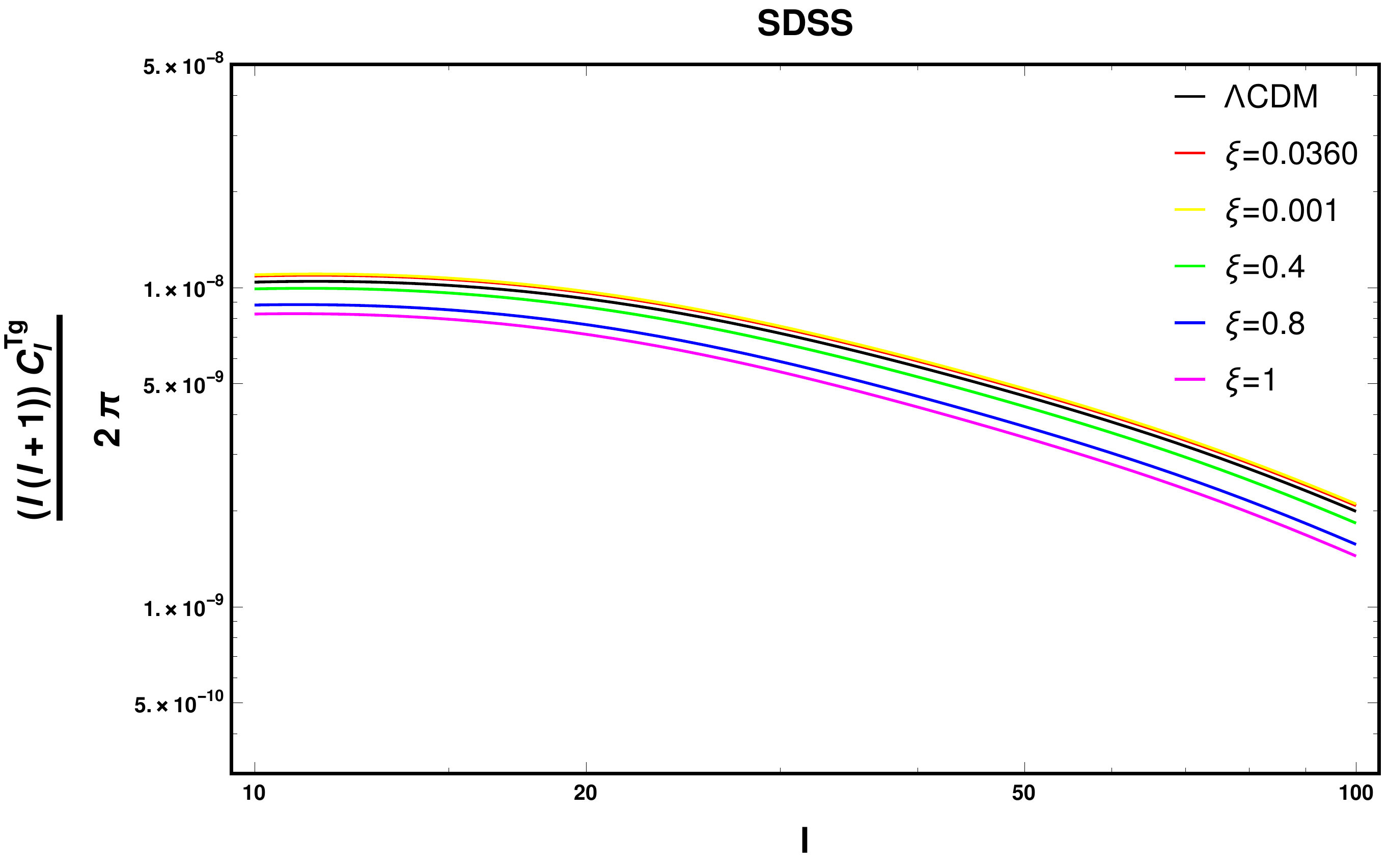}
\caption{The amplitude of the ISW cross power spectrum as a function of multipole order $l$ for the IDMDE and $\Lambda$CDM models. The SDSS survey has been considered. {{\bf{Top:}} For different values of the dark energy equation of state $\omega_{\rm x}$}. {\bf{Bottom:}} For different values of the coupling parameter $\xi$.}
	\label{CTg-SDSS-w}
\end{figure}

\begin{figure}[t!]
\includegraphics[width=\columnwidth]{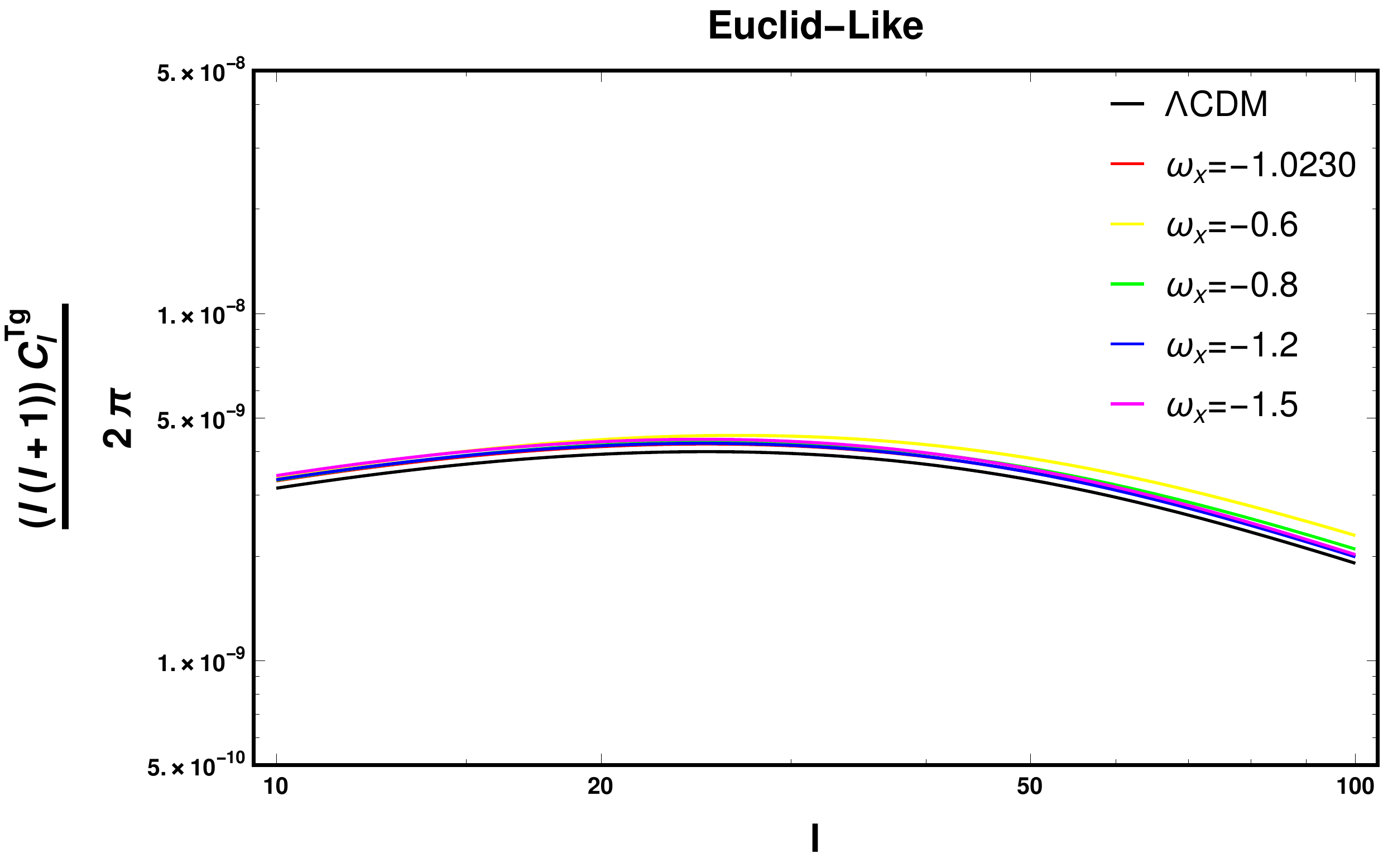}
\\~\\
\includegraphics[width=\columnwidth]{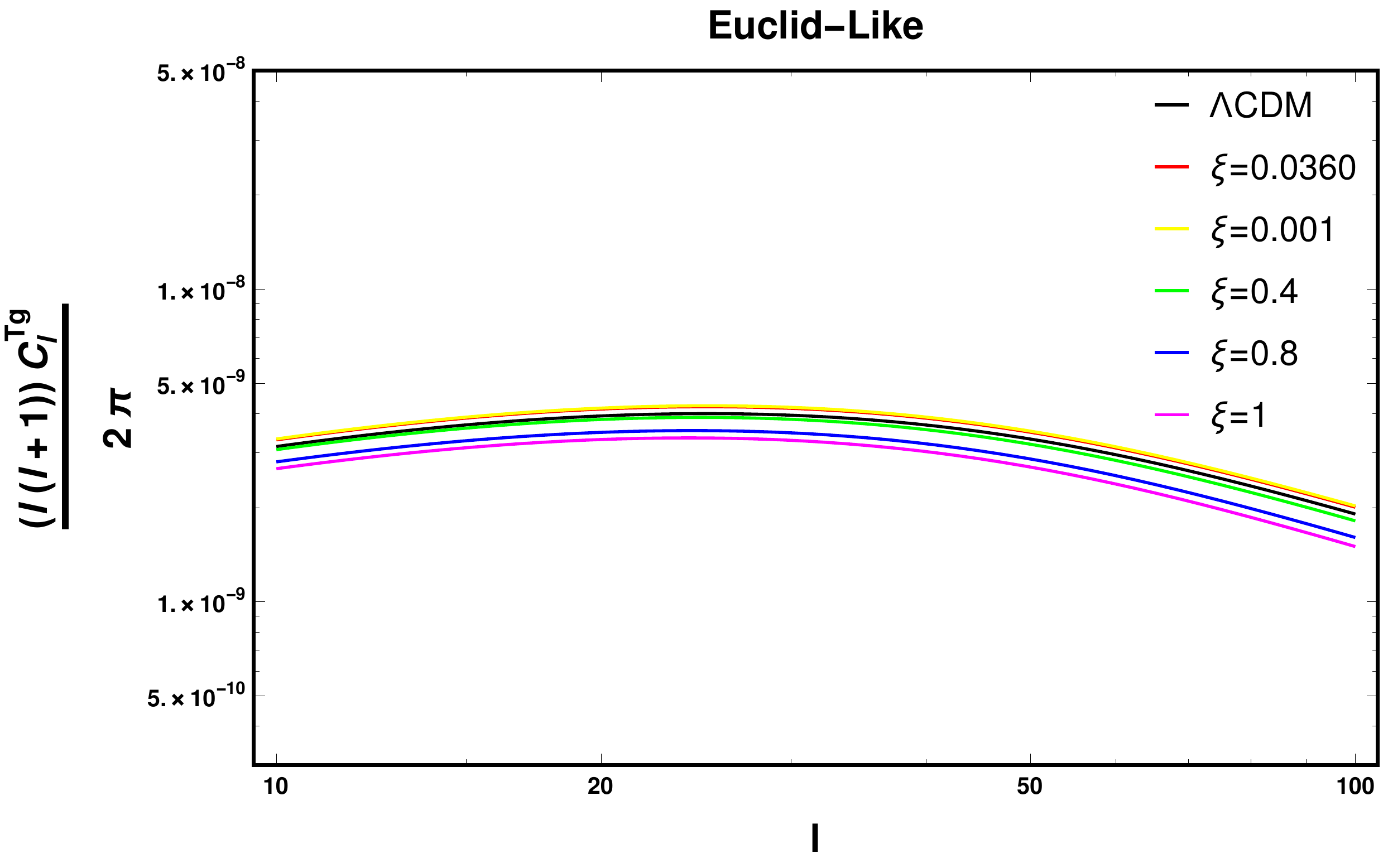}
\caption{The amplitude of the ISW cross power spectrum as a function of multipole order $l$ for the IDMDE and  $\Lambda$CDM models. The Euclid-like survey has been considered. {{\bf{Top:}} For different values of the dark energy equation of state $\omega_{\rm x}$}. {\bf{Bottom:}} For different values of the coupling parameter $\xi$.}
	\label{CTg-Euclid-w}
\end{figure}

We have also shown the corresponding results of other surveys, i.e., the NVSS, SDSS, and Euclid-like, in Figs.~\ref{CTg-NVSS-w}, \ref{CTg-SDSS-w}, and \ref{CTg-Euclid-w}. Similar to what is obtained for the DUNE survey, the results demonstrate that the amplitude of the ISW-cross power spectrum, while considering the IDMDE model, for all values of $\omega_{\rm x}$ is higher than the one obtained from the $\Lambda$CDM model. In addition, for the NVSS, SDSS, and Euclid-like surveys, the results confirm the inverse proportion of the amplitude of the ISW-cross power spectrum with the value of the coupling parameter $\xi$. As a consequence, the IDMDE model with larger coupling parameters leads to smaller amplitudes than the one obtained from the $\Lambda$CDM model.

\section{Conclusions}\label{sec:iv}
In this work, we have calculated the ISW effect in the framework of the IDMDE model, and have compared it with the corresponding result obtained from the $\Lambda$CDM model. The ISW effect is sensitive to the time evolution of the gravitational potentials sourced by large-scale structures. Thus, such an effect seems to be significant in the late-time Universe, which is driven via an accelerated expansion phase. The ISW signal is caused by a stretching effect that results from an imbalance between the structure growth and the cosmic expansion. The rise and fall of large-scale gravitational potentials along the paths of CMB photons in late-time Universe leaves tiny secondary CMB temperature anisotropies on the primary CMB temperature fluctuations. Therefore, CMB photons can be studied to determine the cosmic acceleration and the structure growth rate. In other words, the ISW effect can be used to probe the physics governing the late-time evolution of the Universe and distinguish between possible mechanisms generating cosmic acceleration, including dark energy and modified gravity. 

In this regard, we have initially introduced a theoretical framework of IDMDE models. Additionally, we have determined density and velocity perturbation equations, while considering a perturbed metric for the inhomogeneous Universe. Moreover, we have discussed the stability condition of the IDMDE model. Due to the coupling term in dark energy pressure perturbations, early time instabilities at large scales play a crucial role in the IDMDE model. Hence, the non-adiabatic part of the pressure perturbation might grow quickly at earlier times due to the energy transfer of the IDMDE model, leading to the rapid growth of curvature perturbation on large scales. The results exhibit that by restricting the coupling parameter ($\xi > 0$), large-scale perturbation instabilities can be mitigated for a given energy density transfer rate.

In Fig.~\ref{MPS}, we have shown the amplitude of the matter power spectrum for the IDMDE model, and have compared it with the one obtained from the $\Lambda$CDM model. The results illustrate that the amplitude of the matter power spectrum in the IDMDE model deviates slightly from the one obtained from the $\Lambda$CDM model. In other words, deviations from the $\Lambda $CDM model become larger at lower values of $k$, though they are still much smaller than a half order of magnitude.

Moreover, in Fig.~\ref{F-fig}, we have indicated the evolution of the ISW source term, $\mathcal{F}(a)$, and its derivative with respect to the scale factor in the late-time Universe for the IDMDE model, and have compared it with that of the $\Lambda$CDM model. It is known that the ISW signals passing through the matter-dominated era would be negligible, since the gravitational potential is static during this era. On the other hand, observational data indicate that the Universe appears to be very close to the flat geometry. Thus, as illustrated in Fig.~\ref{F-fig}, the ISW effect can only be explained by the transition to a dark energy-dominated era in the late-time Universe. Also, the results indicate that for the phantom dark energy (i.e., for $\omega_{\rm x}<-1$), the values of $\mathcal{F}(a)$ in IDMDE model is lower than that of the $\Lambda$CDM model at at higher redshifts, whereas the quintessence dark energy (i.e., $\omega_{\rm x}>-1$) in the IDMDE model leads to higher values of $\mathcal{F}(a)$ than the one obtained from the $\Lambda$CDM model. In addition, the derivative of the ISW source term in the IDMDE model is similar to the corresponding result obtained from the $\Lambda$CDM model at higher redshifts, while it deviates from the $\Lambda$CDM model at lower redshifts. Additionally, in the corresponding calculation according to coupling parameter $\xi$, it has been confirmed that for all values of $\xi$, the value of the ISW source term in the IDMDE model is lower than the result obtained for the $\Lambda$CDM model.

In Fig.~\ref{red-dist}, we have shown the normalized redshift distributions of the DUNE, NVSS, SDSS, and Euclid-like surveys as a function of redshift. This distributions span a wide range of redshift up to $z \sim 4$. The results indicate that the NVSS survey leads to the widest redshift coverage, while the SDSS survey yields to the narrowest redshift coverage. Furthermore, the DUNE and Euclid-like surveys provide an intermediate range of redshifts coverage.

In the following, we have calculated the ISW power spectrum within the context of the IDMDE model, and have comapred it with the one obtained from the $\Lambda$CDM model. It is known that the auto-correlation power spectrum and the cross-correlation power spectrum between the CMB temperature and large-scale structures such as galaxies and quasars can be represented in the ISW effect. In this regard, the auto-correlation contribution is subdominant compared to the primordial contributions. Therefore, it cannot be detected via observational data. However, the cross-correlation effect is large enough in a way that it can be detected by multiple studies~\cite{Enander:2015vja}.

In Fig.~\ref{CTT-w}, we have indicated the amplitude of the ISW auto-power spectrum as a function of multipole order $l$ for the IDMDE model, and have comapred it with that obtained from the $\Lambda$CDM model. The results show that for different phantom dark energy equations of state, the amplitude of the ISW auto-power spectrum in the IDMDE model behaves similar to the one for the $\Lambda$CDM model, while, for the quintessence dark energy equations of state, the amplitude of the ISW-auto power spectrum for the IDMDE model is higher than that obtained from the $\Lambda$CDM model. Also, the corresponding results by different values of the coupling parameter demonstrate that $\xi$ is inversely proportional to the amplitude of the ISW-auto power spectrum in the framework of the IDMDE model.

Finally, by employing four different surveys, i.e., the DUNE, NVSS, SDSS, and Euclid-like surveys, we have calculated the amplitude of the ISW-cross power spectrum as a function of multipole order $l$ for the IDMDE model, and have compared them with the corresponding result obtained from the $\Lambda$CDM model. We have also shown the results of these surveys in Figs. \ref{CTg-DUNE-w}, \ref{CTg-NVSS-w}, \ref{CTg-SDSS-w}, and \ref{CTg-Euclid-w}. The results indicate that the amplitude of the ISW-cross power spectrum for the IDMDE model for all values of $\omega_{\rm x}$ is higher than the one obtained for the $\Lambda$CDM model, while deviations are still less than $0.1$ order of magnitude. Moreover, for the case of quintessence dark energy, the amplitude of the ISW-cross power spectrum in the IDMDE model can be in the maximum state. Also, it turns out that the amplitude of the ISW-cross power spectrum in the IDMDE model changes inversely with the value of coupling parameter $\xi$. In addition, it is confirmed that the amplitude of the ISW-cross power spectrum in the IDMDE model for large values of the coupling parameter is lower than that obtained from the $\Lambda$CDM model.

\section*{Acknowledgment}
M.G. and H.F. gratefully thank Abdolali Banihashemi for the useful discussions. Also, M.G. would like to appreciate Bjoern Malte Schaefer and Andras Kovacs for their constructive comments.


\end{document}